\documentclass[12pt,a4paper]{article}
\usepackage{amsfonts,amssymb,amsmath,amsthm,graphicx,bbm,psfrag}
\graphicspath{{images/}} 
\usepackage[english]{babel}
\numberwithin{equation}{section}
\newcommand{\Z}{\mathbb{Z}}
\newcommand{\R}{\mathbb{R}}
\newcommand{\C}{\mathbb{C}}
\newcommand{\dk}[1]{\ddritta \! #1}
\newcommand{\Dt}[2]{\frac{\ddritta\! #1}{\ddritta\! #2}}
\newcommand{\Dtt}[2]{\frac{\ddritta^2 \! #1}{\ddritta\! #2^2}}
\newcommand{\Dttt}[2]{\frac{\ddritta^3 \!#1}{\ddritta\! #2^3}}
\newcommand{\Or}[1]{\mathcal{O}\left(#1\right)}
\newcommand{\e}{\varepsilon}
\newcommand{\wt}{\widetilde}
\newcommand{\abs}[1]{\left\arrowvert #1 \right\arrowvert}
\newcommand{\ff}[1]{\langle#1\rangle}
\newcommand{\Tr}[1]{\Trace\left(#1\right)}
\newcommand{\Det}[1]{\Determinant\left(#1\right)}
\newcommand{\Ai}[1]{\Airy\left(#1\right)}
\newcommand{\E}[1]{\mathbbm{E}\left(#1\right)}
\newcommand{\Pb}[1]{\mathbbm{P}\left(#1\right)}
\newcommand{\ET}[1]{\mathbbm{E}_T\left(#1\right)}
\newcommand{\PbT}[1]{\mathbbm{P}_T\left(#1\right)}

\renewcommand{\phi}{\varphi}
\renewcommand{\Re}{\Reale}
\renewcommand{\Im}{\Immaginario}

\newtheorem{thm}{Theorem}[section]

\newtheorem{prop}[thm]{Proposition}
\newtheorem{lem}[thm]{Lemma}

\DeclareMathOperator*{\ddritta}{d}
\DeclareMathOperator*{\Airy}{Ai}
\DeclareMathOperator*{\Airyp}{Ai^\prime}
\DeclareMathOperator*{\Trace}{Tr}
\DeclareMathOperator*{\Determinant}{Det}
\DeclareMathOperator*{\prodtime}{\prod\!\phantom{ }^{\mathfrak{t}}}
\DeclareMathOperator*{\sgn}{sgn}

\DeclareMathOperator*{\Reale}{Re}
\DeclareMathOperator*{\Immaginario}{Im}

\title{Step fluctuations for a faceted crystal}
\author{Patrik L. Ferrari and Herbert Spohn\\
{\normalsize Zentrum Mathematik and Physik Department, TU M\"unchen}\\{\normalsize D-85747 Garching, Germany}\\
{\normalsize e-mails:\ {\tt  ferrari@ma.tum.de, spohn@ma.tum.de}}}

\begin{document}

\maketitle
\setcounter{tocdepth}{2}     

\begin{abstract}
A statistical mechanics model for a faceted crystal is the 3D Ising model at zero temperature. It is assumed that in one octant all sites are occupied by atoms, the remaining ones being empty. Allowed atom configurations are such that they can be obtained from the filled octant through successive removals of atoms with breaking of precisely three bonds. If $V$ denotes the number of atoms removed, then the grand canonical Boltzmann weight is $q^V$, $0<q<1$. As shown by Cerf and Kenyon, in the limit $q\to 1$ a deterministic shape is attained, which has the three facets 
$(1 0 0)$, $(0 1 0)$, $(0 0 1)$, and a rounded piece interpolating between them. We analyse the step statistics as $q\to 1$. In the rounded piece it is given by a determinantal process based on the discrete sine-kernel. Exactly at the facet edge, the steps have more space to meander. Their statistics is again determinantal, but this time based on the Airy-kernel. In particular, the border step is well approximated by the Airy process, which has been obtained previously in the context of growth models. Our results are based on the asymptotic analysis for space-time inhomogeneous transfer matrices.
\end{abstract}


\section{Introduction}\label{intro}
\begin{figure}[t!]
\begin{center}
\includegraphics[bb=0 50 600 630,clip,height=8cm]{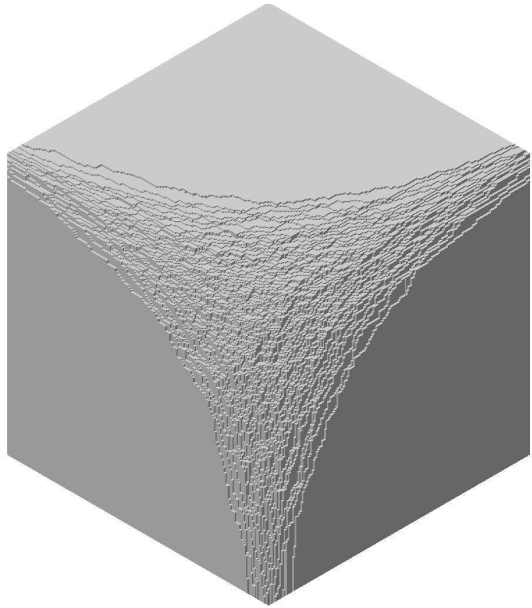}
\caption{\it Crystal corner viewed from the $(1 1 1)$-direction for $q=0.98$.}\label{fig1}
\end{center}
\end{figure}

As a very common phenomenon, crystals are faceted at sufficiently low
temperatures with facets joined through rounded pieces. Of course, on the atomic
scale the crystal surface must be stepped. These steps meander through thermal
fluctuations. On a facet the steps are regularly arranged except for small
errors, whereas on a rounded piece the steps have more freedom to fluctuate. Our
aim is to understand the precise step statistics, where the step bordering the
crystal facet is of particular interest. To gain some insight
we will study a simplified statistical mechanics model of a cubic crystal. Its
equilibrium shape has three facets, each consisting of a part of one of the
coordinate planes. The facets do not touch each other and there is an
interpolating rounded piece, see Figure~\ref{fig1}.
For this model the step statistics will be analyzed in great detail.

Let us first explain our model for the corner of a crystal. The crystal
is assumed to be simple cubic with lattice $\Z^3$.
We use lattice gas language and associate to each site $x\in \Z_+^3$, $\Z_+=\{0,1,2,\ldots\}$,
the occupation variable $n_x=0,1$ with $1$ standing for site $x$ occupied by an atom and $0$
for site $x$ empty.
Up to a chemical potential the binding energy of the
configuration $n$ is

\begin{equation}
H(n)=J \sum_{|x-y|=1}(n_x-n_y)^2,\quad J>0.
\end{equation}
We consider very low temperatures, meaning that all allowed configurations have
the same energy, i.e.\ the same number of broken bonds. To define properly,
we introduce the reference configuration $n^{\mathrm{ref}}$ in which only the octant
$\Z_+^3$ is occupied,
\begin{equation}
n_x^{\mathrm{ref}}=\left\{\begin{array}{cl} 1 & \textrm{ for }x\in\Z_+^3,\\
0 & \textrm{ for }x \in \Z^3 \setminus \Z_+^3.\end{array}\right.
\end{equation}
$n$ is an allowed configuration if for a sufficiently large box $\Lambda$ one
has
\begin{equation}
n_x=n_x^{\mathrm{ref}}\textrm{ for all }x \in \Z_+^3\setminus \Lambda \textrm{
and } H(n)-H(n^{\mathrm{ref}})=0.
\end{equation}
The set of allowed configurations is denoted by $\Omega$.
By construction $\Omega$ is countable. To favor a crystal corner, we introduce the
fugacity $q$, $0<q<1$, and assign to each $n \in \Omega$ the weight
\begin{equation}\label{eqIntro.4}
q^{V(n)},
\end{equation}
where $V(n)$ is the number of atoms removed from $n^\mathrm{ref}$, i.e.
\begin{equation}\label{eqIntro.5b}
V(n)=\sum_{x\in \Z_+^3}(1-n_x).
\end{equation}

A configuration $n \in \Omega$ can uniquely be represented by a height function $h$
over $\Z_+^2$. For the column at $(i,j)\in \Z_+^2$, all sites below $h(i,j)$,
excluding $h(i,j)$, are empty and all sites above $h(i,j)$ are filled.
$n\in\Omega$ if and only if
\begin{equation}\label{eqIntro.5}
h(i+1,j)\leq h(i,j),\quad h(i,j+1)\leq h(i,j),\quad h(i,j)\to 0\textrm{ for } (i,j) \to
\infty.
\end{equation}
By abuse of notation, the set of height functions satisfying (\ref{eqIntro.5}) is also denoted by $\Omega$. For $h\in\Omega$ let $V(h)$ be the volume in $\Z_+^3$ below $h$. Then the weight for the height $h$ is $q^{V(h)}$.

\begin{figure}[t!]
\begin{center}
\psfrag{1}[][][1]{$1$}
\psfrag{2}[][][1]{$2$}
\psfrag{3}[][][1]{$3$}
\includegraphics[height=8cm]{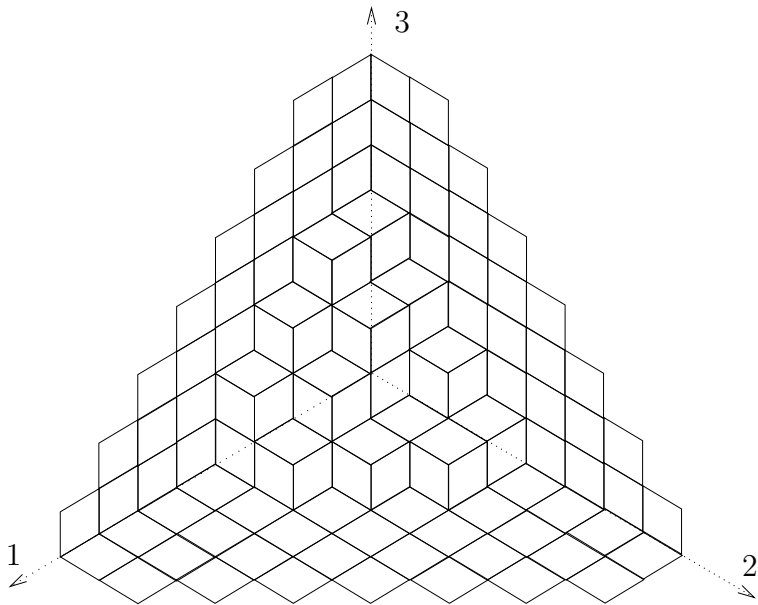}
\caption{\it The $(1 1 1)$ projection of a configuration $n\in\Omega$. In each of
the three sectors the tiling becomes regular far away from the origin.}
\label{lozenge}
\end{center}
\end{figure}
There is an alternative way to describe configurations $n \in \Omega$, which
we just mention for completeness, but will not use later on. One builds the
crystal out of unit cubes and projects its surface along the $(1 1 1)$-direction, which results in a tiling of the plane $\R^2$ with lozenges (rhombi)
oriented along $0$, $2\pi/3$, and $4\pi/3$. With the orientation of Figure~\ref{lozenge}
there are three sectors of the plane corresponding to the polar angle $\theta$
with $-\pi/6 <\theta < \pi/2$, $\pi/2 < \theta < 7\pi/6$, $7\pi/6
< \theta < 11\pi/6$. $n\in\Omega$ if and only if the
tiling in each sector becomes regular sufficiently far away from the origin.

Instead of tilings, if preferred, one can also think of covering the dual
hexagonal lattice by dimers such that every site is covered. In computer science
this is called perfect matching. Equivalently, to have a more statistical
mechanics flavor, one can consider the fully frustrated antiferromagnetic Ising
model on a triangular lattice, i.e\ for an allowed spin configuration every triangle
must have exactly two spins of the same sign. Erasing all bonds connecting equal
sign spins yields a lozenge tiling, and viceversa.

Following the conventional pattern we should now state our main results. This would
mean to bring in a lot of additional notation making our
introduction unwieldy. Thus we sketch only the general goals, explain our
results, and connect to previous studies of the model. The precise theorems
will be given in the respective sections.

The step statistics is studied in the limit $q\to 1$. Thus it is convenient
to set 
\begin{equation}
q=1-\frac{1}{T},\quad T\to \infty.
\end{equation}
Let $h_T$ denote the random height function distributed according to
\begin{equation}\label{eqIntro.8b}
\frac{1}{Z_T}\exp[\ln(1-\tfrac{1}{T})V(h)]
\end{equation} relative to the counting measure on $\Omega$, $Z_T$ the
normalizing partition function. For large $T$ the heights are $\Or{T}$.
Thus one expects a limit shape on scale $T$. In fact, as
proved in~\cite{CK,OR},
\begin{equation}\label{eqIntro.9b}
\lim_{T\to\infty}\frac{1}{T}h_T([uT],[vT])=h_{\mathrm{ma}}(u,v)
\end{equation}
in probability. Here $(u,v)\in \R_+^2$ and $[~~]$ denotes the integer part.
Let $\mathcal{D}=\{(u,v)\in \R_+^2, e^{-u/2}+e^{-v/2} > 1\}.$ On $\mathcal{D}$,
$h_{\mathrm{ma}}$ is strictly decreasing in both coordinates and $h_{\mathrm{ma}}>0$, whereas $h_{\mathrm{ma}}=0$ on $\R_+^2\setminus\mathcal{D}$. The analytic form of $h_{\mathrm{ma}}$ is given in Section~\ref{shape}. If $r$ denotes the distance to $\partial \mathcal{D}= \{(u,v)\in \R_+^2, e^{-u/2}+e^{-v/2} = 1\}$, it follows that $h_{\mathrm{ma}}$ vanishes as $r^{3/2}$.
This is the Pokrovsky-Talapov law.

Our interest here is to zoom to the atomic scale. One possibility is to consider
a macroscopic point $(u,v) \in \cal D$ and the local height statistics
$\{h_T([uT]+i,[vT]+j),(i,j)\in \Z^2\}$. In the limit $T\to\infty$, locally the height profile is planar and one expects that the height statistics corresponds to a random tiling of the plane with the three types of lozenges from Figure~\ref{lozenge},
such that the relative fraction of lozenges yields the average slope $\nabla h_{\mathrm{ma}}(u,v)$.
This property will be proven in Section~\ref{bulk} and we refer to it
as \emph{local equilibrium}: as $T\to\infty$, locally one sees an infinite
volume translation invariant, spatially ergodic Gibbs measure for the lozenges
with their chemical potentials determined through $\nabla h_{\mathrm{ma}}(u,v)$.

\begin{figure}[t!]
\begin{center}
\includegraphics[width=12cm]{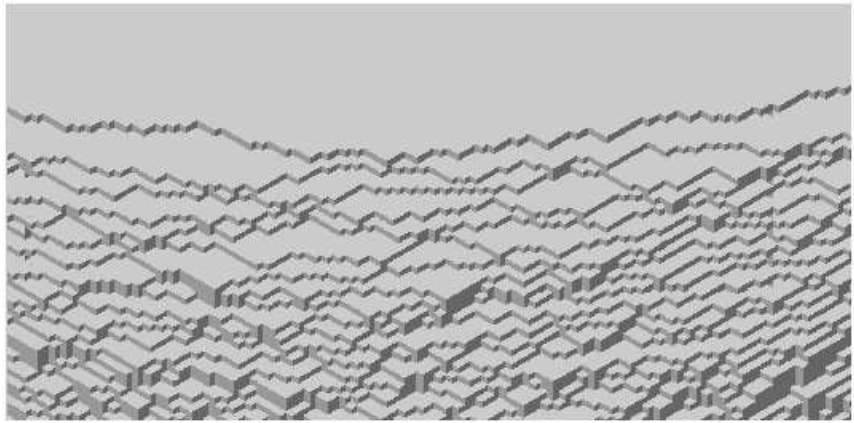}
\caption{\it Zoom to the facet edge in Figure~\ref{fig1}.}\label{figborder}
\end{center}
\end{figure}
An even more intriguing issue is to zoom to the facet edge, which means to take
$(u,v)\in \partial \cal D$, see Figure~\ref{figborder}. Since the step density vanishes at $\partial\cal D$, typically there will be only a few steps in focus. Thus it is more natural to consider directly the crystal step bordering the facet. By symmetry we can choose the border step lying in the $2-3$ plane. Then the border step is given as the graph of the function
\begin{equation}\label{eqIntro.9}
t\mapsto b_T(t)=h_T(0,t),\quad t \in \Z_+.
\end{equation}
From (\ref{eqIntro.5}) we have $b_T(t+1)\leq b_T(t)$ and $\lim_{t\to\infty}b_T(t)=0.$
For large $T$, $b_T$ is $\Or{T}$, and there is a limiting shape according to
\begin{equation}
\lim_{T\to\infty}T^{-1}b_T([\tau T])=b_\infty(\tau),
\end{equation}
where
\begin{equation}\label{eqIntro.11}
b_\infty(\tau)=-2\ln(1-e^{-\tau/2}),\quad \tau>0.
\end{equation}

(\ref{eqIntro.11}) tells us only the rough location of the step. For the step statistics the relevant quantity is the size of the fluctuations of $b_T([\tau T])-T b_{\infty}(\tau)$. As will be shown they are of order $T^{1/3}$ which is very different from steps inside the rounded piece of the crystal which are allowed to fluctuate only as $\ln{T}$. On a more refined level one would like to understand correlations, e.g.\ the joint height statistics at two points $t$ and $t'$. They have a systematic part corresponding to $b_\infty(\tau)$. Relative to it the correlation length along the border step scales as $T^{2/3}$, which reflects that on short distances the border step looks like a Brownian motion. Thus $b_\infty$ has to be expanded including the curvature term and the correct scaling for the
border step is
\begin{equation}\label{eqIntro.12}
T^{-1/3}\!\Big\{b_T([\tau T\!+\!s T^{2/3}])\!-\!\left(b_\infty(\tau) T\! +\!b^\prime_\infty(\tau) s T^{2/3} \!+\!\tfrac{1}{2} b_\infty^{\prime\prime}(\tau) s^2  T^{1/3}\right)\Big\}\!=\! A_T(s).
\end{equation}
Here $\tau>0$ is a fixed macroscopic reference point and $s\in \R$ with $s T^{2/3}$
the longitudinal distance. $s\mapsto A_T(s)$ is regarded as a stochastic
process in $s$. Our central result is the convergence
\begin{equation}
\lim_{T\to\infty}A_T(s)=\kappa A(s \kappa/2)
\end{equation}
in the sense of convergence of finite dimensional distributions.
The limit process $A(s)$ is the stationary Airy process. Its scale is determined
by the local curvature via $\kappa=\sqrt[3]{2b_\infty^{\prime\prime}(\tau)}$.
The Airy process appeared first in the study of shape fluctuations for the
polynuclear growth model~\cite{PS}. It can also be obtained through edge scaling
of $\beta=2$, GUE Dyson's Brownian motion.

\section{Line ensemble, determinantal process}\label{lines}
\subsection{Gradient lines}
\begin{figure}[t!]
\begin{center}
\psfrag{1}[][][1]{$1$}
\psfrag{2}[][][1]{$2$}
\psfrag{3}[][][1]{$3$}
\includegraphics[height=6cm]{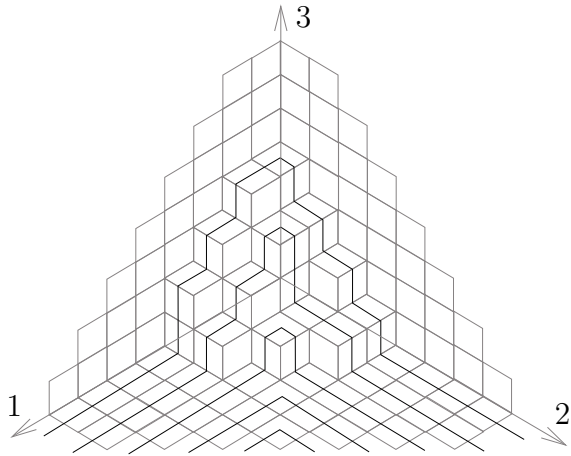}
\caption{\it The gradient lines for the tiling of Figure~\ref{lozenge}.}
\label{gradient}
\end{center}
\end{figure}
In view of Figure~\ref{lozenge}, it is natural to represent $h$ in term of its
level lines with the hope that they have a tractable statistics.
In fact, it turns out to be more convenient to consider the gradient lines as
drawn in Figure~\ref{gradient}.
\begin{figure}[t]
\begin{center}
\psfrag{t}[][][1]{$t$}
\psfrag{j}[][][1]{$j$}
\psfrag{1}[][][1]{$1$}
\psfrag{h0}[][][1]{$h_0$}
\psfrag{h1}[][][1]{$h_1$}
\psfrag{h2}[][][1]{$h_2$}
\psfrag{h3}[][][1]{$h_3$}
\includegraphics[width=6cm]{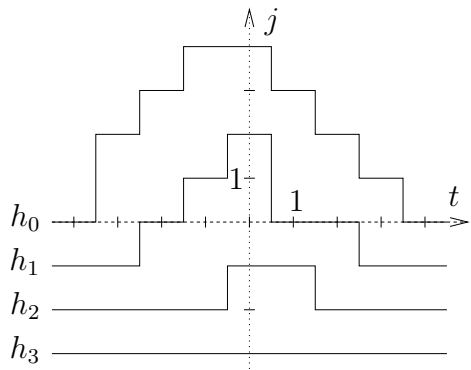}
\caption{\it The gradient lines for the tiling of Figure~\ref{lozenge}.}\label{LineEnsemble}
\end{center}
\end{figure}
In Figure~\ref{LineEnsemble} the underlying lattice is distorted in such a way
that the gradient lines become ``trajectories'' on a square lattice. It is this
latter representation which will be used in the sequel. Clearly, the surface
statistics can be reconstructed from the statistics of the line ensemble. As
first noticed by Okounkov and Reshetikhin~\cite{OR} the occupation number field
corresponding to the line ensemble of Figure~\ref{LineEnsemble} has
determinantal correlations. In this section we will rederive their results using
the fermionic framework, which is a convenient starting point for our asymptotic
analysis.

The gradient lines of Figures~\ref{gradient},~\ref{LineEnsemble} are defined
through
\begin{equation}\label{eqLine.1}
t=j-i, \quad h_\ell(t)=h(i,j)-\ell(i,j),
\end{equation}
where
\begin{equation}
\ell(i,j)=(i+j-|i-j|)/2
\end{equation}
labels the line, $(i,j)\in\Z_+^2$. $h_\ell$ is increasing for $t\leq 0$ and decreasing
for $t\geq 0$,
\begin{equation}\label{eqLine.2}
h_\ell(t)\leq h_\ell(t+1), \quad t\leq 0,\qquad h_\ell(t)\geq h_\ell(t+1), \quad t\geq 0,
\end{equation}
with the asymptotic condition
\begin{equation}\label{eqLine.3b}
\lim_{t\to\pm \infty} h_\ell(t)=-\ell.
\end{equation}
By construction the gradient lines satisfy the non-crossing constraint
\begin{equation}\label{eqLine.4b}
h_{\ell+1}(t)<h_{\ell}(t-1), \quad t\leq 0, \qquad 
h_{\ell+1}(t)<h_{\ell}(t+1), \quad t\geq 0.
\end{equation}
Height configurations $h\in \Omega$ are mapped one-to-one to gradient lines
satisfying (\ref{eqLine.2}), (\ref{eqLine.3b}), and (\ref{eqLine.4b}).

We extend $h_\ell$ to piecewise constant functions on $\R$ such that the jumps are at midpoints, i.e.\ at points of $\Z+\tfrac{1}{2}$. For a given line, $h_\ell$, let $t_{\ell,1}<\ldots < t_{\ell,k(\ell)}<0$ be the left jump times with jump heights $s_{\ell,1},\ldots,s_{\ell,k(\ell)}$ and let $0<t_{\ell,k(\ell)+1}<\ldots <t_{\ell,k(\ell)+n(\ell)}$ be the right jump times with jump heights $-s_{\ell,k(\ell)+1},\ldots,-s_{\ell,k(\ell)+n(\ell)}$. It follows from (\ref{eqIntro.4}), (\ref{eqIntro.5b}) that the weight for the line configuration $\{h_\ell\}_{\ell=0,1,\ldots}$ is given by 
\begin{equation}\label{eqLine.3}
\prod_{\ell=0}^{\infty}\exp\bigg[\ln(1-1/T) \bigg(\sum_{j=1}^{k(\ell)+n(\ell)} s_{\ell,j} |t_{\ell,j}|\bigg)\bigg].
\end{equation}

The line ensemble with weight (\ref{eqLine.3}) can be thought of as world lines of
``fermions'', where $t$ refers to time and $j$ to space. It
is then natural to introduce the random field of occupation variables,
denoted by $\eta(j,t)$.
Thus $\eta(j,t)=1$ if there is a line passing at $(j,t)$ and $\eta(j,t)=0$
otherwise. As to be shown, the random field $\eta$ has determinantal correlation
functions, which as one crucial ingredient relies on the non-crossing
constraint (\ref{eqLine.4b}). However, in previous applications only nearest
neighbor jumps appear, whereas our model has the unusual feature that jumps of
arbitrary size are allowed.

\subsection{Fermions}
The basic tool is the transfer matrix from $t$ to $t+1$, $t \in \Z$.
A fermion is created (resp.\ annihilated) at the position $j \in \Z$ by the operator $a^*_j$ (resp.\ $a_j$). The CAR algebra $\{a^*_j,a_j, j \in \Z\}$ over $\Z$ is defined by the anticommutation relations
\begin{equation}\label{eqLine.4}
\{a_i,a_j\}=0,\quad \{a^*_i,a^*_j\}=0,\quad \{a_i,a^*_j\}=\delta_{i,j}
\end{equation}
for $i,j\in \Z$.
First we consider $t\leq -1$, in which case only up-steps can occur.
To each unit up-step at time $t$ we assign the weight $q_t=q^{\abs{t+1/2}}$ which satisfy (\ref{eqLine.3}).
The rule is that in a jump from $i$ to $j$, $j\geq i$, one creates additional
particles at sites $m$ with $i+1\leq m \leq j$ and annihilates particles at
sites $m$ with $i\leq m \leq j-1$.
E.g.\ if a fermionic world line jumps from $-1$ to $3$,
one creates particles at positions $0,1,2,3$, and annihilates
the particles at $-1,0,1,2$. This rule ensures the non-crossing constraint
(\ref{eqLine.4b}), since, if two fermionic world lines would intersect, 
a fermion is created twice at the same position, which leads to a
zero contribution. The corresponding rule applies to $t\geq 0$ with the
difference that the jumps are downwards only.

Let us define the operators 
\begin{equation}
b_l=\sum_{k\in \Z}{a^*_{k+l} a_k}.
\end{equation}
The transfer matrix from $t$ to $t+1$ is a sum of the $n$-step transitions $T_n$ as
\begin{equation}
\widehat{T}(t,t+1)=\mathbbm{1}+q_t T_1+q_t^2 T_2+\ldots + q_t^n T_n + \ldots,
\end{equation}
where 
\begin{equation}
T_n=\frac{(-1)^n}{n!}\sum_{k_1,\ldots,k_n}{a_{k_1}\ldots a_{k_n} a^*_{k_n+1}
\ldots a^*_{k_1 +1}}.
\end{equation}
The $(-1)^n$ prefactor results from the left ordering of the $a_j$ and $a^*_j$'s.

We would like to reexpress $T_n$  in terms of products of
the $b_i$'s only. For $n,m>0$ the commutators are
\begin{equation}
b_n a^*_k = a^*_k b_n+a^*_{k+n},\quad b_n a_k = a_k b_n-a_{k-n},
\quad [b_n,b_m]=0, \quad [b_{-n},b_{-m}]=0.
\end{equation}
These relations lead to
\begin{equation}
T_n=\sum_{\begin{subarray}{c}d_1,\ldots,d_n\geq 1\\d_1+2d_2+\ldots =
n\end{subarray}}{\prod_{j=1}^{n}{\left(\frac{b_j}{j}\right)^{d_j}\frac{1}{d_j!}}}.
\end{equation}
The Schur polynomials $\{p_k(y)\}_{k\geq 0}$ are polynomials such that
\begin{equation}\label{eqLine.10}
\exp\bigg(\sum_{k\geq 1}{t^k y_k}\bigg)= \sum_{l\geq 0}{p_l(y) t^l},
\quad y=y_1,y_2,\ldots,
\end{equation}
and given explicitly by
\begin{equation}
p_l(y)=\sum_{\begin{subarray}{c}x_1,\ldots,x_l\geq 1\\
x_1+2x_2+\ldots = l\end{subarray}}{\prod_{j=1}^{l}\frac{y_j^{x_j}}{x_j!}}.
\end{equation}
Comparing with (\ref{eqLine.10}) yields
\begin{equation}
\widehat{T}(t,t+1)=\sum_{l\geq 0}{q_t^l T_l}= \exp\bigg(\sum_{k\geq 1}{q_t^k
\frac{b_k}{k}}\bigg).
\end{equation}
We conclude that the transfer matrix is given by
\begin{equation}\label{eqLine.13}
\widehat{T}(t,t+1)=\exp\bigg(\sum_{k\geq 1}{q^{k\abs{t+1/2}}
\frac{b_k}{k}}\bigg)
\end{equation}
for $t \in \Z_-=\{-1,-2,\ldots\}$, and, by the same reasoning,
\begin{equation}\label{eqLine.13b}
\widehat{T}(t,t+1)=\exp\bigg(\sum_{k\geq 1}{q^{k (t+1/2)} \frac{b_{-k}}{k}}\bigg)
\end{equation}
for $t\in \Z_+$.

$\widehat{T}(t,t+1)$ is quadratic in the fermion operators.
Hence it is the second quantization of a one-particle operator acting of
$\ell_2$. For easier reading second quantization is merely indicated by a
``$\widehat{\phantom{m}}$'', i.e.\ for $A$ acting of $\ell_2$ we set $\widehat A=\Gamma(A)$ as its
second quantization. From (\ref{eqLine.13}), (\ref{eqLine.13b}) we read off
\begin{equation}
T(t,t+1)=\exp\bigg(\sum_{k\geq 1} \frac{q^{k\abs{t+1/2}}}{k} \alpha_k\bigg)
\end{equation}
for $t \in \Z_-$, and
\begin{equation}
T(t,t+1)=\exp\bigg(\sum_{k\geq 1} \frac{q^{k (t+1/2)}}{k} \alpha_{-k}\bigg)
\end{equation}
for $t\in \Z_+$ with matrices $\alpha_k$ defined through
\begin{equation}
[\alpha_k]_{i,j}=\left\{\begin{array}{ll}1 &\textrm{ if }i-j=k,\\
0& \textrm{ otherwise.}\end{array}\right.
\end{equation}
$T(t,t+1)$ are invertible with the $\ell_2$-norms
\begin{eqnarray}\label{eqLine.21a}
\|T(t,t+1)\|&\leq& \exp\left(\frac{q^{|t+1/2|}}{1-q^{|t+1/2|}}\right) ,\nonumber \\
\| T(t,t+1)^{-1}\| &\leq& \exp\left(\frac{q^{|t+1/2|}}{1-q^{|t+1/2|}}\right) .
\end{eqnarray}

For the state at $t=\pm \infty$ all sites in $\Z_-\cup\{0\}$ are filled, those
in $\Z_+\setminus \{0\}$ are empty, which, together with the transfer matrices
(\ref{eqLine.13}), (\ref{eqLine.13b}) determines the Green's functions of an imaginary time
(Euclidean) Fermi field. It is inhomogeneous in space-time and uniquely
given through the two-point function $\ff{a_i^*(t)a_j(t')}$.
To  compute it correctly one has to employ the standard finite volume
approximation. We first restrict all world lines to lie in the spatial interval $[-M,M]$.
Thereby the transfer matrix depends on $M$ in the sense that all creation and
annihilation operators with index $\abs{k} > M$ are set equal to zero. The state
at $\pm\infty$ is $(1,\ldots,1,0,\ldots,0)^t$ which is $2M+1$ long with the last
$1$ at site $0$. The projector on this state is approximated through
\begin{equation}
\exp[\beta \widehat N_M]
\end{equation}
in the limit $\beta\to\infty$ with
$\widehat N_M=\sum_{k=-M}^0a_k^*a_k-\sum_{k=1}^M a_k^*a_k$.
We first compute the equal time Green's function through
\begin{eqnarray}\label{eqLine.15}
& &\hspace{-20pt}\ff{a^*_i(t_0) a_j(t_0)}_T =\\
&=& \hspace{-6pt}\lim_{M\to\infty}\lim_{L\to\infty}\lim_{\beta\to\infty}
\frac{1}{Z_{\beta,M,L}}{\Tr{e^{\beta \widehat N_M}\prodtime_{t=t_0}^{L}
\widehat T_M(t,t+1) a^*_i a_j \prodtime_{t=-L}^{t_0-1}\widehat T_M(t,t+1)}},
\nonumber
\end{eqnarray}
where the trace is over the antisymmetric Fock space $\cal F$ with one-particle
space $\ell_2([-M,\ldots,M])$. The products are time-ordered increasingly from
right to left, which is indicated by the superscript $\mathfrak{t}$ at the product symbol $\prod$. $Z_{\beta,M,L}$ is the normalizing partition function, which is
defined through the same trace with $a_i^*a_j$ replaced by $\mathbbm{1}$. As
explained in Appendix~\ref{app1}, (\ref{eqLine.15}) can be expressed in terms of
one-particle operators as the limit $M,L,\beta\to\infty$ of
\begin{equation}\label{eqLine.16a}
\ff{a^*_i(t_0) a_j(t_0)}_{T,\beta M L} =
\left[\mathbbm{1}+\left(\prodtime_{t=-L}^{t_0-1}
T_M(t,t+1)e^{\beta N_M} \prodtime_{t=t_0}^{L}T_M(t,t+1)\right)^{-1}\right]^{-1}_{j,i}.
\end{equation}

Let $P_+ + P_- = \mathbbm{1}$  in $\ell_2$ with $P_+$ the projection onto
$\Z_+\setminus\{0\}$ and let
\begin{equation}
e^{G_{\textrm{right}}(t_0)}=\prod_{t=t_0}^{\infty}T(t,t+1), \quad
e^{G_{\textrm{left}}(t_0)}=\prod_{t=-\infty}^{t_0-1}T(t,t+1),
\end{equation}
and
\begin{equation}
e^{G_{\uparrow}(t_0)}=\prod_{t=-\infty}^{\min(0,t_0)-1}T(t,t+1), \quad
e^{G_{\downarrow}(t_0)}=\prod_{t=\max(0,t_0)}^{\infty}T(t,t+1).
\end{equation}
By (\ref{eqLine.21a}) the infinite products are well-defined, as are their
inverses. The $T(t,t+1)$'s commute and no time-ordering is required. Hence
\begin{eqnarray}\label{Gpm}
G_{\uparrow}(t_0) &=& \sum_{t=-\infty}^{\min(0,t_0)-1}
\sum_{r\geq 1}{\frac{q^{r\abs{t+1/2}}}{r}\alpha_r}=
\sum_{r\geq 1}\mu_r(t_0)\alpha_r,\nonumber\\
G_{\downarrow}(t_0) &=& \sum_{t=\max(0,t_0)}^{\infty}\sum_{r\geq 1}
\frac{q^{r(t+1/2)}}{r}\alpha_{-r}=\sum_{r\geq 1}{\nu_r(t_0) \alpha_{-r}}
\end{eqnarray}
with \begin{equation}
\mu_r(t_0)=\frac{q^{r/2}q^{-r\min(0,t_0)}}{r(1-q^r)}, \quad
\nu_r(t_0)=\frac{q^{r/2}q^{r\max(0,t_0)}}{r(1-q^r)}.
\end{equation}
In (\ref{eqLine.16a}) we take limits as indicated in (\ref{eqLine.15}). Then
\begin{equation}\label{eqLine.17a}
\ff{a^*_i(t_0) a_j(t_0)}_T =
\left[e^{G_{\textrm{left}}(t_0)} P_- (P_- e^{G_{\textrm{right}}(t_0)}
e^{G_{\textrm{left}}(t_0)} P_-+P_+)^{-1} P_- e^{G_{\textrm{right}}(t_0)}\right]_{j,i}.
\end{equation}

Let \begin{equation}
e^{G_{-}}=\prod_{t=-\infty}^{-1}T(t,t+1), \quad
e^{G_{+}}=\prod_{t=0}^{\infty}T(t,t+1).
\end{equation}
Then $e^{G_{\textrm{right}}(t_0)}e^{G_{\textrm{left}}(t_0)}=e^{G_{+}}e^{G_{-}}=e^{G_{-}}e^{G_{+}}$
and, decomposing $\ell_2=P_-\ell_2\oplus P_+\ell_2$, we have
\begin{equation}
e^{G_{-}}=\left[\begin{array}{cc}a&0\\c&b\end{array}\right],\quad
e^{G_{+}}=\left[\begin{array}{cc}a' & c' \\ 0 &b'\end{array}\right].
\end{equation}
Thus
\begin{equation}
P_- (P_- e^{G_{\textrm{right}}(t_0)}e^{G_{\textrm{left}}(t_0)} P_-+P_+)^{-1} P_- =
\left[\begin{array}{cc}(a a')^{-1} & 0 \\ 0 & 0\end{array}\right]
\end{equation}
and, since
\begin{equation}
e^{-G_{-}}=\left[\begin{array}{cc}a^{-1}& 0 \\ -b^{-1} c a^{-1} &
b^{-1}\end{array}\right],\quad
e^{-G_{+}}=\left[\begin{array}{cc}(a')^{-1} & - (a')^{-1}c' (b')^{-1} \\ 0 &
(b')^{-1}\end{array}\right],
\end{equation}
we obtain
\begin{equation}
e^{-G_{+}} P_- e^{-G_{-}} =P_- (P_- e^{G_{\textrm{right}}(t_0)}
e^{G_{\textrm{left}}(t_0)} P_-+P_+)^{-1} P_-.
\end{equation}
Therefore
\begin{equation}
\ff{a^*_i(t_0) a_j(t_0)}_T =\left[e^{G_{\textrm{left}}(t_0)}e^{-G_{+}} P_- e^{-G_{-}} e^{G_{\textrm{right}}(t_0)}\right]_{j,i},
\end{equation}
which rewrites as
\begin{equation}\label{eqLine.17}
\ff{a^*_i(t_0) a_j(t_0)}_T =
\left[ e^{G_{\uparrow}(t_0) - G_{\downarrow}(t_0)}
P_- e^{-(G_{\uparrow}(t_0) - G_{\downarrow}(t_0))}\right]_{j,i}.
\end{equation}
The Fermi field depends on $T$ through $q=1-1/T$. For this reason we keep the
index $T$. 

Using the anticommutation relations (\ref{eqLine.4}) in (\ref{eqLine.15}) we
immediately obtain
\begin{equation}\label{eqLine.19}
\ff{a_j(t_0) a^*_i(t_0)}_T =\left[ e^{G_{\uparrow}(t_0) - G_{\downarrow}(t_0)}
P_+ e^{-(G_{\uparrow}(t_0) - G_{\downarrow}(t_0))}\right]_{j,i}.
\end{equation}
Thus our final result for the equal time correlations reads
\begin{eqnarray}\label{eqLine.20}
\ff{a_i^*(t_0)a_j(t_0)}_T&=&\sum_{l\leq 0}
{\big[e^{G_{\uparrow}(t_0)-G_{\downarrow}(t_0)}\big]_{j,l}
\big[e^{-G_{\uparrow}(t_0)+G_{\downarrow}(t_0)}\big]_{l,i}},\nonumber\\
\ff{a_j(t_0)a_i^*(t_0)}_T&=&\sum_{l>0}
{\big[e^{G_{\uparrow}(t_0)-G_{\downarrow}(t_0)}\big]_{j,l}
\big[e^{-G_{\uparrow}(t_0)+G_{\downarrow}(t_0)}\big]_{l,i}}.
\end{eqnarray}

To extend (\ref{eqLine.20}) to unequal times we have to go through the same
limit procedure as before. Since the argument is in essence unchanged,
there is no need to repeat. We define the propagator from  $a$ to $b$, $a\leq b$, through
\begin{equation}
e^{G(a,b)}=\prod_{t=a}^{b-1}T(t,t+1),\quad e^{G(a,a)}=\mathbbm{1}, \quad e^{G(b,a)}=e^{-G(a,b)}.
\end{equation}
Using the identity
\begin{equation}\label{OperEvol}
e^{-G(0,t_0)} a_{m} e^{G(0,t_0)}= \sum_{k\in\Z}{\big[e^{G(0,t_0)}\big]_{m,k}a_{k}}
\end{equation}
for $t\geq t'$, the full two-point function is given by
\begin{eqnarray}\label{eqLine.22}
\ff{a_{j}^*(t)a_{j'}(t')}_T&=&\sum_{l\leq 0}
{\big[e^{G_{\uparrow}(0)-G_{\downarrow}(0)+G(0,t')}}\big]_{j',l}
\big[e^{-G_{\uparrow}(0)+G_{\downarrow}(0)-G(0,t)}\big]_{l,j},\nonumber \\
\ff{a_{j}(t)a_{j'}^*(t')}_T &=&
\sum_{l>0}{\big[e^{G_{\uparrow}(0)-G_{\downarrow}(0)+G(0,t)}}\big]_{j,l}
\big[e^{-G_{\uparrow}(0)+G_{\downarrow}(0)-G(0,t')}\big]_{l,j'}. \qquad
\end{eqnarray}

\subsection{Determinantal random field $\eta(j,t)$}
Moments of the random field $\eta(j,t)$ introduced above can be expressed
through fermionic correlations. We consider first equal time correlations. The
basic identity is
\begin{equation}\label{eqLine.37}
\ET{\prod_{k=1}^{n}\eta_T(j_k,t)}=\bigg\langle
\prod_{k=1}^{n}a_{j_k}^*(t)a_{j_k}(t)\bigg\rangle_T,
\end{equation}
where $\mathbb{E}_T$ is the expectation with respect to the normalized weight
(\ref{eqLine.3}). If $\{j_1,\ldots,j_n\}$ are distinct, then, as explained in Appendix~\ref{app2}, the fermionic expectation is determinantal and
\begin{equation}\label{DetEq}
\ET{\prod_{k=1}^n \eta(j_k,t)}=\Det{R_T(j_k,t; j_l,t)}_{1\leq k, l \leq n},
\end{equation}
with
\begin{equation}\label{eqLine.39}
R_T(i,t;j,t)=\ff{a^*_{i}(t)a_{j}(t)}_T.
\end{equation}
If coinciding arguments are admitted, then (\ref{DetEq}) still holds with the convention
\begin{equation}\label{DetEq2}
R_T(i,t;j,t)=\left\{\begin{array}{ll}\ff{a^*_{i}(t)a_{j}(t)}_T
& \textrm{ for }i \leq j,\\
\ff{a^*_{i}(t)a_{j}(t)}_T-\delta_{i,j}=-\ff{a_{j}(t)a^*_{i}(t)}_T
&\textrm{ for }i > j.\end{array}\right.
\end{equation}

(\ref{eqLine.37}) is easily extended to unequal time correlations. Let us
consider $n$ disjoint space-time points $(j_1,t_1),\ldots,(j_n,t_n)$ ordered
increasingly as $t_1\leq t_2 \leq \ldots \leq t_n$. Then the basic identity is
\begin{equation}\label{eqLine.27a}
\ET{\prod_{k=1}^n{\eta(j_k,t_k)}} = \ff{a^*_{j_n}(t_n)a_{j_n}(t_n)\cdots
a^*_{j_1}(t_1)a_{j_1}(t_1)}_T.
\end{equation}
Using (\ref{OperEvol}) the left hand side equals
\begin{equation}\label{eqLine.27b}
\sum_{\begin{subarray}{c}k_1,\ldots,k_n\\l_1,\ldots,l_n \end{subarray}}
{\prod_{q=1}^n \big[e^{-G(0,t_q)}\big]_{k_q,j_q}
\big[e^{G(0,t_q)}\big]_{j_q,l_q} \ff{a^*_{k_n}a_{l_n}\cdots a^*_{k_1}a_{l_1}}_T}.
\end{equation}
Let us set
\begin{equation}\label{eqLine.28}
R_T(j,t; j',t')=\left\{\begin{array}{l@{\textrm{ for }}l} \ff{a^*_{j}(t) a_{j'}(t')}_T
&t \geq t', \\ -\ff{a_{j'}(t')a^*_{j}(t)}_T & t < t'.\end{array}\right.
\end{equation}
Then the unequal time correlations are given by
\begin{equation}\label{eqLine.27c}
\ET{\prod_{k=1}^n{\eta(j_k,t_k)}}=\Det{R_T(j_k,t_k; j_l,t_l)}_{1\leq k,l \leq n}.
\end{equation}

The identity (\ref{eqLine.27c}) has been derived from left to right. One can
read it also from right to left. Then $R_T$ is the defining kernel, resp.\ 
Green's function, which is considered to be given and (\ref{eqLine.27c}) defines
the moments of some determinantal space-time random field over $\Z\times\Z$. Of
course, $R_T$ cannot be chosen arbitrarily, since the right hand side of
(\ref{eqLine.27c}) must be moments of a probability measure. For determinantal
random fields over the space coordinate only, compare with (\ref{DetEq}), proper conditions on the defining kernel have been studied in detail~\cite{Soshnikov, Takahashi}. The space-time variant is less well understood, see \cite{Jo:DetProc} for a discussion.

The determinantal property is preserved under limits. Thus through bulk and edge scaling further determinantal space-time random fields will be encountered below. One of them is over $\Z\times\Z$ with equal-time given through the sine-kernel. The other is over $\Z\times\R$ with equal-time given through the Airy kernel.

\section{Limit shape}\label{shape}
On the macroscopic scale, in the limit $T\to\infty$, the random field
$\eta(j,t)$ becomes deterministic with a profile given by
\begin{equation}\label{eqShape.1}
\rho(\zeta,\tau)=\left\{\begin{array}{l@{\textrm{ for }}l}1& \zeta \leq
b_\infty^-(\tau),\\ \frac{1}{\pi}\arccos\left(\cosh(\tau/2)-e^{-\zeta+\abs{\tau}/2}/2\right)
& b_\infty^-(\tau)<\zeta<b_\infty(\tau),\\ 0 & \zeta\geq b_\infty(\tau),\end{array}\right.
\end{equation}
with
\begin{equation}
b_\infty^-(\tau)=-2\ln(1+e^{-\abs{\tau}/2}),\quad b_\infty(\tau)=-2\ln(1-e^{-\abs{\tau}/2}).
\end{equation}
More precisely, for all continuous test functions $f:\R^2\to \R$ with compact
support
\begin{equation}\label{eqShape.3}
\lim_{T\to\infty}\frac{1}{T^2}\sum_{j,t}f(j/T,t/T)\eta(j,t) =
\int\dk{\zeta}\dk{\tau}\rho(\zeta,\tau)f(\zeta,\tau)
\end{equation}
almost surely. (\ref{eqShape.3}) assumes more spatial averaging than needed. In
fact, it suffices to choose a test function whose support on the scale of
the lattice diverges as $T\to\infty$ and to properly normalize.

As a consequence of (\ref{eqShape.3}) the limit (\ref{eqIntro.9b}) holds. $h_{\textrm{ma}}$ can be read off from (\ref{eqShape.1}) and is given in parametric form through
\begin{equation}\label{eqShape.4}
h_{\textrm{ma}}(u,v)=\left\{\begin{array}{l@{\textrm{ for }}l}
0 & (u,v)\in \R_+^2\setminus \mathcal{D},\\
\frac{1}{2}(u+v-\abs{\tau})+ \zeta(u,v) & (u,v)\in \mathcal{D},
\end{array}\right.
\end{equation}
where $\tau=v-u$ and where $\zeta(u,v)$ is the unique solution $\zeta$  of the equation
\begin{equation}
\frac{1}{2}(u+v-\abs{\tau})=\int_{b_\infty^-(\tau)}^\zeta
(1-\rho(\zeta',\tau))\dk{\zeta'}-\zeta
\end{equation}
in the interval $[b_\infty^-(\tau),b_\infty(\tau)]$.
While the limit (\ref{eqShape.3}) has been established by Okounkov and Reshetikhin~\cite{OR}, compare also with Section~\ref{bulk}, the existence of the limit shape has been
proved before by Cerf and Kenyon~\cite{CK}. Instead of (\ref{eqIntro.8b}), they
used the fixed volume constraint $V(h)= 2\zeta_R(3) T^3$, resp.\ $V(h)\leq
2\zeta_R(3) T^3$, with $\zeta_R$ the Riemann zeta function. They write the limit shape $S_0$, as a set of $\R^3$, in the parametric representation
\begin{equation}\label{eqShape.6}
S_0=\{2(f(a,b,c)-\ln{a},f(a,b,c)-\ln{b},f(a,b,c)-\ln{c}) \, | \, a,b,c >0\}
\end{equation}
with
\begin{equation}\label{eqShape.7}
f(a,b,c)=\frac{1}{4\pi^2}\int_{0}^{2\pi}\dk{u}\int_{0}^{2\pi}\dk{v}
\ln(a+b e^{iu}+c e^{iv}).
\end{equation}
Here $a,b,c$ denote the weights for the three orientations of the lozenges and
$f(a,b,c)$ is the corresponding free energy per unit area for the lozenge tiling of
the plane. As expected from the equivalence of ensembles, the shapes given by
(\ref{eqShape.4}) and (\ref{eqShape.6}) are identical. This can be seen as follows. Let $z=(z_1,z_2,z_3)$ represent a point on the limit shape. We compare
$z_2-z_1$ and $z_3-z_1$ (resp.\ $z_3-z_2$) for $z_2\geq z_1$ (resp.\ 
$z_2\leq z_1$) for the parametrizations (\ref{eqShape.4}) and (\ref{eqShape.6}).
This leads to $a=1$, $b=e^{-\tau/2}$, $c=e^{-\zeta/2}$ for $z_2\geq z_1$
and to $b=1$, $a=e^{-\abs{\tau}/2}$, $c=e^{-\zeta/2}$ for $z_2\leq z_1$. Since
(\ref{eqShape.7}) is symmetric in $a,b,c$, one verifies that indeed
\begin{equation}
\int_{-2\ln(1+e^{-\abs{\tau}/2})}^\zeta
\hspace{-1.6cm}(1-\rho(\zeta',\tau))\dk{\zeta'}=
2 f(1,e^{-\abs{\tau}/2},e^{-\zeta/2})+\zeta.
\end{equation}

According to (\ref{eqShape.4}), $h_{\textrm{ma}}=0$ on $\R_+^2\setminus \mathcal{D}$. Close to the edge the height vanishes with the power $3/2$. E.g.\ in the direction $\tau=v-u$ one has \begin{equation}
h_{\textrm{ma}}(r,\tau)\simeq\cosh(\tau/4)\pi^{-1} 2^{1/4}r^{3/2}
\end{equation}
with $r$ the distance to the edge. The $3/2$ power is known as Pokrovsky-Talapov law~\cite{PT}.

A limit shape theorem is a law of large numbers. It is available also for related tiling models. A famous case is the Aztec diamond~\cite{CEP}. Cohn, Larsen and Propp \cite{CLP} consider the 3D-Young diagrams constrained to the box $\alpha N\times \beta N\times \gamma N$ with $\alpha,\beta,\gamma \sim \Or{1}$ and compute the limit shape as $N\to\infty$. In the line-ensemble representation their model corresponds to $q=1$ with the boundary conditions that at $t=-\alpha N,\beta N$ all lattices sites are occupied expect for those in the interval $[1,\gamma N]$. \cite{CLP} computed the line density and from it the limit shape. Two or higher order point functions are not studied. From our representation we see that higher order correlation functions are determinantal even in this case. However the computation of the two-point function is more complicated, since one cannot rely any more on an expression like (\ref{eqLine.17}). For a list of further limit shape theorems we refer to the survey~\cite{K1}.

The limit shape can be determined through minimizing the appropriate macroscopic free energy functional. The input is the microscopic surface tension at given slope $\nabla h$. For example in the $(1 1 1)$-frame the surface tension $\sigma_{(111)}(\nabla h)$ is given by (\ref{eqShape.7}), where $a,b,c$ are defined through the prescribed surface tilt 
$\nabla h$. $\sigma_{(111)}$ has been computed in ~\cite{Kas,Wu,BH}. Correspondingly there is a surface tension in the $(0 0 1)$-frame, denoted by $\sigma_{(001)}(\nabla h)$. 

To obtain the free energy $\mathcal{F}$ for some macroscopic height profile $h$ over a bounded domain $\mathcal{B}$, one argues that $h$ is made up of little planar pieces, each one of them having the surface tension at the corresponding local slope. Adding up yields
\begin{equation}\label{eq1.8}
\mathcal{F}(h)=\int_{\mathcal{B}} \dk{u}\dk{v} \, \sigma_{(001)}(\nabla h(u,v)).
\end{equation}
In our case we have $\mathcal{B}=\R_+^2$, $h$ is decreasing in both variables
such that $h(u,v)=0$ for $(u,v)\to\infty$, and $V(h)=\int_{\mathcal{B}}\dk{u}\dk{v}\,h(u,v)$. The minimizer of $\mathcal{F}$, under these constraints and $V(h)=2\zeta_R(3)$, is $h_{\textrm{ma}}$ from (\ref{eqShape.4}). Equivalently one could minimize $\mathcal{F}(h)+V(h)$.

Probabilistically, $\mathcal{F}(h)+V(h)$ can be viewed as a large deviation functional in the sense that in the limit $T\to\infty$, with respect to the normalized probability $Z^{-1}q^{V(h)}$,
\begin{equation}\label{eqIntro.11c}
\PbT{T^{-1}h_T([uT],[vT])\simeq h}=\Or{e^{-T^2 (\mathcal{F}(h)+V(h) -\mathcal{F}(h_{\textrm{ma}})-V(h_{\textrm{ma}}))}}
\end{equation}
for given macroscopic height profile $h$~\cite{CK}.

Expanding (\ref{eqIntro.11c}) to quadratic order in $\delta h=h-h_{\textrm{ma}}$ yields a heuristic formula for the covariance of the Gaussian shape fluctuations. In spirit it is proportional to $(-\partial_u^2-\partial_v^2)^{-1}$, hence like a massless Gaussian field. This implies in particular, that on the macroscopic scale shape fluctuations are small, of order $\ln{T}$ only. Gaussian fluctuations are proved for the Aztec diamond in~\cite{Jo:RandTilings} and for domino tilings of a Temperleyan polyomino in~\cite{K:Gaussian}.

The limit shape theorem (\ref{eqShape.1}) implies that also the border step has
a deterministic limit. We state a result, which is stronger than what could be deduced from (\ref{eqShape.1}) and which follows by the transfer matrix techniques to be explained in Section~\ref{edge}.

\begin{thm} Let $b_T$ be the border step as defined in (\ref{eqIntro.9}). Then
for any $\delta>0$, $c>0$, $0<u_-<u_+<\infty$ one has
\begin{equation}
\lim_{T\to\infty}\Pb{|T^{-1} b_T([u T])-b_\infty(u)|\geq c T^{-2/3+\delta},
u_-\leq u \leq u_+}=0.
\end{equation}
\end{thm}

\section{Bulk scaling, local equilibrium}\label{bulk}
For local equilibrium we zoom to a point $(\zeta_0,\tau_0)T$ with $b_\infty^-(\tau_0)<\zeta_0<b_\infty(\tau_0)$ at average density $\rho=\rho(\zeta_0,\tau_0)$, which means to consider the random field
\begin{equation}
\eta_{T}^{\mathrm{bulk}}(j,t)=\eta([\zeta_0 T]+j,[\tau_0 T]+t)
\end{equation}
with $(j,t)\in \Z^2$ and $[~~]$ denoting the integer part. Properly speaking we should keep the reference point $(\zeta_0,\tau_0)$ in our notation. Since it is fixed throughout, we suppress it for simplicity. In the limit $T\to\infty$, $\eta_{T}^{\mathrm{bulk}}(j,t)$ becomes stationary. Then at fixed $t$, one has to fill the Fermi states up to the density $\rho$ which implies that $\eta_{\infty}^{\mathrm{bulk}}(j,t)$, $t$ fixed, is a determinantal point process on $\Z$ as defined through the discrete sine-kernel. Only at $\tau_0=0$, the inhomogeneity of the underlying $\eta$-field can still be seen, which, of course, is an artifact of our coordinate system. In the $(1 1 1)$ projection the line $\tau_0=0$ would be just like any other local slope with a corresponding stationary distribution of lozenges. The case $\tau_0=0$ can also be treated. For simplicity we omit it and require $\tau_0\neq 0$.

Let us define the kernel $S(j,t;j',t')$ by
\begin{equation}\label{eqBulk.5}
S(j,t;j',t')=
\frac{\sgn(t-t')}{2\pi}\int_{I(t,t')}\hspace{-0.6cm}\dk{k}
\exp\left[ik(j'-j)+(t'-t)\ln(1-e^{-\abs{\tau_0}/2}e^{-i k})\right]
\end{equation} for $\tau_0>0$ and
\begin{equation}\label{eqBulk.5b}
S(j,t;j',t')=\frac{\sgn(t-t')}{2\pi}\int_{I(t,t')}\hspace{-0.6cm}\dk{k}
\exp\left[ik(j'-j)-(t'-t)\ln(1-e^{-\abs{\tau_0}/2}e^{i k})\right]
\end{equation} for $\tau_0<0$,
where $$I(t,t')=\left\{\begin{array}{l@{\textrm{ if }}l}[-\pi \rho,\pi\rho],
&t\geq t', \\ \phantom{ }[\pi\rho,2\pi-\pi\rho], &t<t',\end{array}\right.$$
and $\sgn(t-t')=1$, if $t\geq t'$, and $\sgn(t-t')=-1$, if $t<t'$.
In particular at equal times
\begin{equation}\label{eqLine.sine}
S(i,t;j,t)=\frac{\sin(\rho \pi (i-j))}{\pi (i-j)},
\end{equation}
which is the sine-kernel. $S$ depends on the reference point $(\zeta_0,\tau_0)$. In the particular case of equal times the dependence is only through the local density.

\begin{thm}\label{BulkThm}
In the sense of convergence of local distributions we have
\begin{equation}
\lim_{T\to\infty}\eta_{T}^{\mathrm{bulk}}(j,t)=\eta^{\mathrm{sine}}(j,t).
\end{equation}
For $\tau_0>0$, $\eta^{\mathrm{sine}}(j,t)$ is the determinantal space-time random field with defining kernel (\ref{eqBulk.5}) and for $\tau_0<0$ the one with the kernel (\ref{eqBulk.5b}).
\end{thm}
\noindent {\bf Remark:} Theorem~\ref{BulkThm} is identical to Theorem 2 of~\cite{OR}. We use here an integral representation for the defining kernel $R_T$ which differs somewhat from the one of~\cite{OR} and which turns out to be convenient in the context of the edge scaling.
\begin{proof}[\sc{Proof:}]
We consider the case $\tau_0 > 0$ only, since $\tau_0<0$ follows by symmetry.
Let us set
\begin{equation}
B_T(j,t;j',t')=e^{g(j)-g(j')}R_T([\zeta_0 T]+j,[\tau_0 T]+t;
[\zeta_0 T]+j',[\tau_0 T]+t'),
\end{equation}
where $R_T$ is defined in (\ref{eqLine.28}) and
$g(j)=j\abs{\tau_0}T\ln(1-1/T)/2$.
The determinant in (\ref{eqLine.27c}) does not change under similarity
transformation, in particular not under multiplying by $e^{g(u_i)-g(u_j)}$. 
Therefore
\begin{equation}
\ET{\prod_{k=1}^{m}\eta_{T}^{\mathrm{bulk}}(j_k,t_k)}=\Det{B_T(j_k,t_k;j_l,t_l)}_{1\leq k,l \leq m}
\end{equation}
and we need to prove that pointwise
\begin{equation}
\lim_{T\to\infty}B_T(j,t;j',t')=S(j,t;j',t').
\end{equation}

First consider $t\geq t'$. For $\tau_0>0$ we take $T$ large enough so that
$\tau_0 T+t'\geq 0$ (this simplifies (\ref{eqBulk.11}) below). Using 
(\ref{eqLine.27c}) we obtain
\begin{eqnarray}\label{eqBulk.9b}
B_T(j,t;j',t') &=& e^{g(j)-g(j')} \\
& \times &\sum_{l\leq 0}\big[e^{G_{\uparrow}(\tau_0 T)-G_{\downarrow}(\tau_0 T+t)}\big]_{\zeta_0 T+j,l}
\big[e^{G_{\downarrow}(\tau_0 T+t')-G_{\uparrow}(\tau_0 T)}\big]_{l,\zeta_0 T+j'}.
\nonumber
\end{eqnarray}
An explicit expression for the matrix elements of the two-point
functions can be found using the translation invariance of the one-particle
operators. In Fourier representation they are given by
\begin{equation}\label{eqLine.23}
\left[\exp\bigg(\sum_{r\in\Z}\sigma_r \alpha_r\bigg)\right]_{n,m}=
\frac{1}{2\pi}\int_{-\pi}^{\pi}\exp(-i k (n-m)) \exp\bigg(\sum_{r\in\Z}\sigma_r
e^{i k r}\bigg)\dk{k}
\end{equation}
for $\sigma_r \in \R$.
Then using (\ref{eqLine.23}) and changing $l$ into $-l$, we have
\begin{eqnarray}\label{eqBulk.9}
B_T(j,t;j',t') &= &\sum_{l\geq 0}
\frac{e^{g(j)}}{2\pi}\int_{-\pi}^{\pi}{\dk{k} e^{\sigma(k)T}e^{\phi_q(k,t)}
e^{-i k (\zeta_0 T+j)}e^{-i k l}} \nonumber \\
&\times&\frac{e^{-g(j')}}{2\pi}\int_{-\pi}^{\pi}{\dk{k'} e^{-\sigma(k')T}e^{-\phi_q(k',t')}
e^{i k' (\zeta_0 T+j')}e^{i k' l}}, 
\end{eqnarray}
where
\begin{equation}
\sigma(k)=(1-q)\sum_{r\geq 1}{\frac{q^{r/2}}{r(1-q^r)}
(e^{ikr}-q^{r \tau_0 /(1-q)}e^{-ikr})}
\end{equation}
and
\begin{equation}\label{eqBulk.11}
\phi_q(k,t)=
\sum_{r\geq 1}{\frac{q^{r/2} (1-q^{r t})}{r (1-q^r)}q^{r\tau_0/(1-q)} e^{-i k r}}.
\end{equation}

To study the asymptotic of integrals as (\ref{eqBulk.9}) we consider the complex $k$ plane and regard the integration in (\ref{eqBulk.9}) as being along the real line from $-\pi$ to $\pi$. Such a line integral can be deformed to another path $C$ with the same endpoints. The complex integration along $C$ will be denoted by $\int_{C}\dk{k}\cdots$. In the particular case when the path is on the real line, say from $a$ to $b$, the integral will be denoted by $\int_{a}^{b}\dk{k}\cdots$.

Let us consider the following four paths: $\xi_0=-\pi\to\pi$, $\xi_1(p)=-\pi+i
p\to \pi+i p$, $\xi_2=-\pi\to-\pi+i p$, and $\xi_3=\pi+i p\to \pi$ with $0\leq
p\leq \tau_0$. The factors in (\ref{eqBulk.9}) are integrals along $\xi_0$.
Their integration contour can be deformed from $\xi_0$ to $\xi_2\circ \xi_1\circ \xi_3$ without changing the integrals, since the integrands are holomorphic.
Moreover the integrals on $\xi_2$ and $\xi_3$ cancel exactly because of 
periodicity of the integrands. We transform the integral in $k$ into the
integral along $\xi_1(\theta)$ and the one in $k'$ into the integral
along $\xi_1(\theta+\e)$, with $0<\e\ll 1$ and
$\theta=-\tau_0 T \ln(1-1/T)/2$. $\theta$ is chosen such that the exponentially
large function in $T$ passes through the critical point of $\sigma(k')$.
Consequently we have
\begin{eqnarray}\label{eqBulk.12}
B_T(j,t;j',t') &=& \frac{e^{g(j)-g(j')}}{(2\pi)^2}  \int_{\xi_1(\theta)}\dk{k'}\int_{\xi_1(\theta+\e)}
\dk{k}e^{(\sigma(k)-\sigma(k'))T}e^{i\zeta_0 T(k'-k)} \nonumber\\
& &\hspace{1.2cm}\times \,\,e^{\phi_q(k,t)-\phi_q(k',t')} e^{i(k'j'-kj)} (1-e^{i(k'-k)})^{-1}.
\end{eqnarray}
As $T\to\infty$ we obtain
\begin{equation}
\sigma(k)=2 i\sum_{r\geq 1}\frac{e^{-r\tau_0/2}}{r^2}\sin((k-i\tau_0/2)r)+\Or{1/T}.
\end{equation}
Therefore the terms that increase or decrease exponentially in $T$ in (\ref{eqBulk.12}) are $E(k)$ and $-E(k')$, where
\begin{equation}
E(k)=2 i\sum_{r\geq 1}\frac{e^{-r\tau_0/2}}{r^2}\sin((k-i\tau_0/2)r)-i\zeta_0 k.
\end{equation}
The critical points of $E(k)$ are
\begin{equation}
\pm k_c+i\tau_0/2,\quad
k_c=\arccos\left(\cosh(\tau_0/2)-\frac{e^{-\zeta_0+\tau_0/2}}{2}\right)\in \R.
\end{equation}
For $\Im(k)=\tau_0/2$, $\Re(E(k))=\zeta_0 \tau_0/2$,
the analysis of $\Re(E(k))$ for $k$ close to the line $\Im(k)=\tau_0/2$ shows
that, for $\Re(E(k))\in[-\pi,-k_c]\cup [k_c,\pi]$, it decreases when increasing
$\Im(k)$ and, for $\Re(E(k))\in [-k_c,k_c]$, it decreases when decreasing $\Im(k)$.

Next we transform the integral into a sum of three terms, the first two vanish
as $T\to\infty$ and the third one gives the final result, see Figure~\ref{integrals}.
We have $\int_{I_0}\dk{k'}\dk{k} \cdots =\int_{I_1}\dk{k'}\dk{k} \cdots 
+\int_{I_2}\dk{k'}\dk{k} \cdots +\int_{I_3}\dk{k'}\dk{k} \cdots$, where the
integrand is the one of (\ref{eqBulk.12}). Let us compute the three integrals
separately.
\begin{figure}[t!]
\begin{center}
\psfrag{piu}[][][1.5]{$+$}
\psfrag{Re}[][][0.8]{$\Re$}
\psfrag{Im}[][][0.8]{$\Im$}
\psfrag{I0}[][][0.8]{$I_0$}
\psfrag{I1}[][][0.8]{$I_1$}
\psfrag{I2}[][][0.8]{$I_2$}
\psfrag{I3}[][][0.8]{$I_3$}
\psfrag{kk}[][][0.8]{$k$}
\psfrag{k}[][][0.8]{$k'$}
\psfrag{e}[][][0.8]{$\tilde{\e}$}
\includegraphics[width=14cm]{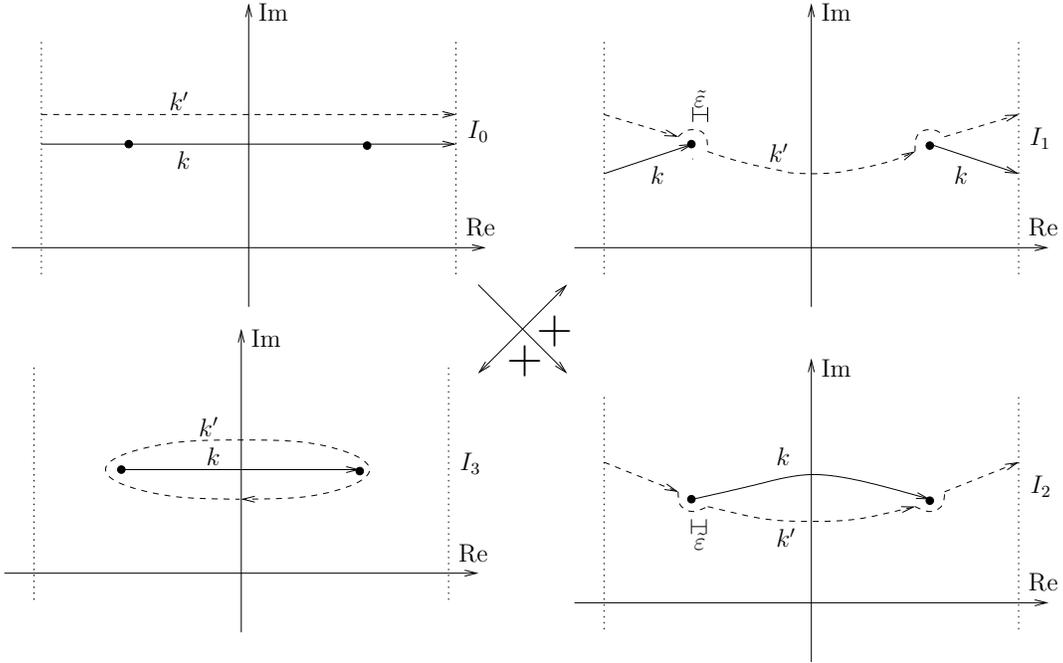}
\caption{\it Deformation of integration paths. The original integral, along $I_0$,
is deformed to the sum of integrals along $I_1$, $I_2$, and $I_3$.
$k$ is integrated along the dashed lines and $k'$ along the solid lines.
The full dots are the critical points of $\sigma(k)$.}\label{integrals}
\end{center}
\end{figure}
For the integration along $I_3$, first we integrate out $k$ taking
the residuum at $k=k'$. Then changing the variable to $z=k'-i\theta$ we obtain
\begin{equation}
\int_{I_3}{\dk{k'}\dk{k}\cdots} = \frac{1}{2\pi}\int_{-k_c}^{k_c}\dk{z}
e^{\phi_q(z+i\theta,t)-\phi_q(z+i\theta,t')} e^{iz(j'-j)}.
\end{equation}
The asymptotic of $\phi_q(z+i\theta,t)$ is
\begin{equation}\label{eqBulk.17}
\phi(z,t)=\lim_{q\to 1}\phi_q(z+i\theta,t)= -t\ln(1-e^{-\tau_0/2}e^{-i z}).
\end{equation}
The integrals along $I_1$ and $I_2$ are treated in the same way. Let us
estimate, e.g., the one along $I_1$. First we integrate in $k$.
The integral $\abs{\int\dk{k}\cdots}$, such that the integration avoids
the two arcs of circle of radius $\tilde \e$ around the critical points (see
Figure~\ref{integrals}), is bounded by $\Or{e^{-a \tilde{\e} T}/(\tilde{\e} T)}$
for some $a>0$. $\Or{e^{-a \tilde{\e} T}/T}$ comes from integrating
$e^{E(k) T}$ and $\Or{1/\tilde{\e}}$ because the minimum of $\abs{k-k'}$ equals
$\tilde{\e}$. The integration through the two arcs around the critical
points is bounded by $\Or{e^{a' \tilde{\e} T}/(\tilde{\e} T)}$ for some $a'>0$, because the integrand is at most of $\Or{e^{a' \tilde{\e} T}/\tilde{\e}}$ for some $a'>0$ and the length of the path of integration is $\Or{1/T}$. We choose therefore $\tilde{\e}=1/T$, so that $\abs{\int\dk{k}\cdots}\leq \Or{1}$. The integration in $k'$ gives an extra-factor $\Or{1/T}$, and
\begin{equation}
\lim_{T\to\infty}\int_{I_1}{\dk{k'}\dk{k}\cdots}=0.
\end{equation}

Summarizing for $t\geq t'$, we have proved that, 
\begin{equation}\label{eqBulk.19}
\lim_{T\to\infty}{B_T(j,t;j',t')}= \frac{1}{2\pi}\int_{-\rho \pi}^{\rho
\pi}\dk{z} e^{\phi(z,t-t')}e^{i z(j'-j)},
\end{equation}
where $\rho=k_c/\pi$ and $\phi(z,t)$ as given in (\ref{eqBulk.17}).
In particular for $t=t'$, $\phi(z,t-t')=0$, which implies (\ref{eqLine.sine}).
The case $t<t'$ is treated in a similar way, leading to
\begin{equation}\label{eqBulk.20}
\lim_{T\to\infty}{B_T(j,t;j',t')}= -\frac{1}{2\pi}\int_{\rho \pi}^{2\pi-\rho
\pi}\dk{z} e^{\phi(z,t-t')}e^{i z(j'-j)}.
\end{equation}
Therefore
\begin{eqnarray}\label{eqBulk.23}
\lim_{T\to\infty}\ET{\prod_{k=1}^m\eta_{T}^{\mathrm{bulk}}(j_k,t_k)}
&=& \Det{S(j_k,t_k;j_l,t_l)}_{1\leq k,l \leq m} \nonumber\\
&=& \mathbb{E}_{\mathrm{b}}\left(\prod_{k=1}^m \eta^{\mathrm{sine}}(j_k,t_k) \right). 
\end{eqnarray}

The proof for $\tau_0<0$ is identical.
\end{proof}

(\ref{eqBulk.19}), (\ref{eqBulk.20}) define a space-time homogeneous Fermi field. Physically it corresponds to fermions on the lattice $\Z$ in their ground state at density $\rho$ and with kinetic energy
\begin{equation}\label{EqBulk.24}
E_{\textrm{kin}}(k)=\ln(1-e^{-\abs{\tau_0}/2-i k \sgn{\tau_0}}).
\end{equation}
$E_{\textrm{kin}}$ is complex reflecting that the fermions have a drift.

The moments (\ref{eqBulk.23}) define a probability measure $\mathbb{P}_{\mathrm{b}}$ on the lozenge tilings of the plane, where the relative fraction of their type depends on the reference point $(\zeta_0,\tau_0)$. $\mathbb{P}_{\mathrm{b}}$ is a Gibbs measure in the sense that its conditional expectations satisfy the DLR equations. We refer to~\cite{FS} of how DLR equations are adjusted in the context of surface models. $\mathbb{P}_{\mathrm{b}}$ is translation invariant with a definite fraction of each type of lozenges. $\mathbb{P}_{\mathrm{b}}$ is even spatially mixing, since truncated correlations decay to zero. One would expect that $\mathbb{P}_{\mathrm{b}}$ is the unique Gibbs measure with these properties. A proof would require that the same limit measure $\mathbb{P}_{\mathrm{b}}$ is obtained when other boundary conditions are
imposed, at fixed lozenge chemical potentials. To our knowledge, only for the surface model studied in \cite{FS} such a uniqueness property has been established.

\section{Edge scaling}\label{edge}
For the edge scaling one zooms at a macroscopic point lying exactly on the border of the facet, i.e.\ at $(\zeta_0,\tau_0)T$ with $\zeta_0=b_\infty(\tau_0)$. For simplicity we set $\tau_0>0$. $\tau_0<0$ follows by symmetry. Since at the edge the step density is zero, one has to consider a scale coarser than the one for the bulk scaling in Section~\ref{bulk}. From our study of the PNG droplet we know already that the longitudinal scale is $T^{2/3}$ and the transversal scale is $T^{1/3}$. On that scale the curvature of $b_\infty$ cannot be neglected. Therefore the correct reference points are
\begin{eqnarray}
t(s)&=&[\tau_0 T+s T^{2/3}],\\
j(r,s)&=&[b_\infty(\tau_0) T+b_\infty'(\tau_0) s T^{2/3}+
\tfrac{1}{2}b_\infty''(\tau_0) s^2 T^{1/3}+r T^{1/3}]. \nonumber
\end{eqnarray}
Note that $(r,s)\in \R^2$. The discrete lattice disappears under edge scaling. Let
us abbreviate
\begin{equation}
\begin{array}{rcrcl}
\alpha_1&=&b_\infty(\tau_0)&=&-2\ln(1-e^{-\tau_0/2}),\\
\alpha_2&=&-b_\infty'(\tau_0)&=&e^{-\tau_0/2}/(1-e^{-\tau_0/2}),\\
\alpha_3&=&b_\infty''(\tau_0)&=&e^{-\tau_0/2}/2(1-e^{-\tau_0/2})^2. 
\end{array}
\end{equation}
Then the edge-scaled random field reads
\begin{equation}\label{eqEdge.4}
\eta^{\mathrm{edge}}_T(r,s)=T^{1/3}\eta_T([\alpha_1 T-\alpha_2 s T^{2/3}+
\tfrac{1}{2}\alpha_3 s^2 T^{1/3}+r T^{1/3}],[\tau_0 T+s T^{2/3}]).
\end{equation}
The prefactor $T^{1/3}$ is the volume element for $r T^{1/3}$. Properly speaking we should keep the reference time $\tau_0$. Since it is fixed throughout, we suppress it in our notation.

Since $\eta^{\mathrm{edge}}_T$ is determinantal, so must be its limit.
For the PNG droplet under edge scaling the limit is the Airy random field and,
by universality, in our model the steps close to the facet edge should have the
same statistics in the limit $T\to\infty$. The Airy field is determinantal
in space-time with Green's function
\begin{equation}\label{eqBulk.8PS}
K^{\mathrm{Airy}}(r,s;r',s')=\sgn(s'-s)\int_\R\dk{\lambda} \, \theta(\lambda(s-s'))
e^{\lambda(s'-s)} \Ai{r-\lambda}\Ai{r'-\lambda},
\end{equation}
where the step function $\theta(s)=0$, if $s<0$, and $\theta(s)=1$, if $s\geq 0$.
The Airy field is stationary in time. In particular, the equal time correlations are given through the Airy kernel
\begin{eqnarray}
K^{\mathrm{Airy}}(r,s;r',s)&=&\int_{-\infty}^0\dk{\lambda}\Ai{r-\lambda}\Ai{r'-\lambda}
\\ &=&\frac{1}{r-r'}\big(\Airy(r)\Airyp(r')-\Airy(r')\Airyp(r)\big).\nonumber
\end{eqnarray}

\begin{thm}\label{thmEdge}
Under edge scaling (\ref{eqEdge.4}) the correlation functions have the following limit,
\begin{equation}
\lim_{T\to\infty}\ET{\prod_{k=1}^m\eta^{\mathrm{edge}}_T(r_k,s_k)} =
\E{\prod_{k=1}^m \left(\kappa^{-1}\eta^{\mathrm{Airy}}\left(\frac{r_k}{\kappa},
\frac{\kappa}{2}s_k\right)\right)}
\end{equation}
uniformly for $r_k,s_k$ in a bounded set. Here 
$\kappa=\sqrt[3]{2 b_\infty''(\tau_0)}$.
In particular for the process
$\eta^{\mathrm{edge}}_T(f,s)=\int\dk{x}f(x)\eta^{\mathrm{edge}}_T(x,s)$,
smeared over continuous test functions $f:\R\to\R$ with compact support, one has
\begin{equation}
\lim_{T\to\infty}\eta^{\mathrm{edge}}_T(f,s)=\int\dk{x}f(\kappa x) \eta^{\mathrm{Airy}}(x,s \kappa/2)
\end{equation}
in the sense of the convergence of  joint finite-dimensional distributions.
\end{thm}

To prove Theorem~\ref{thmEdge} one only has to establish that under edge scaling (\ref{eqLine.28}) converges to (\ref{eqBulk.8PS}). We define the rescaled kernel (\ref{eqLine.28}) as
\begin{equation}\label{eqEdge7}
K_T(r,s;r',s')=\frac{e^{-g(r,s)}}{e^{-g(r',s')}}
T^{1/3}R_T(j(r,s),t(s);j(r',s'),t(s'))
\end{equation}
where $g(r,s)=-j(r,s)(\tau_0 T \ln(1-1/T)/2+s T^{2/3}\ln(1-1/T)/2)$
and $R_T(j,t;j',t')$ from (\ref{eqLine.28}).

\begin{prop}\label{propEdge}
The edge-scaled kernel (\ref{eqEdge7}) converges to the Airy kernel
\begin{equation}\label{eqEdge.9b}
\lim_{T\to\infty}K_T(r,s;r',s')=\kappa^{-1}K^{\mathrm{Airy}}
\left(\frac{r}{\kappa},\frac{\kappa}{2}s;\frac{r'}{\kappa},
\frac{\kappa}{2}s'\right)
\end{equation}
uniformly for $r,r',s,s'$ in bounded sets.
\end{prop}

Granted Proposition~\ref{propEdge} we establish Theorem~\ref{thmEdge}.
\begin{proof}[\sc{Proof of Theorem~\ref{thmEdge}:}]
From (\ref{eqLine.27c}) and (\ref{eqEdge.4}) it follows that
\begin{equation}
\ET{\prod_{k=1}^m\eta^{\mathrm{edge}}_T(r_k,s_k)}
=\Det{T^{1/3}R_T(j(r_k,s_k),t(s_k);j(r_l,s_l),t(s_l))}_{1\leq k,l\leq m}.
\end{equation}
This determinant does not change when multiplied by the factor
$e^{-g(r,s)+g(r',s')}$ and therefore
\begin{equation}
\ET{\prod_{k=1}^m\eta^{\mathrm{edge}}_T(r_k,s_k)}
=\Det{K_T(r_k,s_k;r_l,s_l)}_{1\leq k,l\leq m}.
\end{equation}
Note that $g(r,s)$ diverges as $T\to\infty$. On the other hand
\begin{equation}
\E{\prod_{k=1}^m \left(\kappa^{-1}\eta^{\mathrm{Airy}}
\left(\frac{r_k}{\kappa},\frac{\kappa}{2}s_k\right)\right)}
=\Det{\kappa^{-1}K^{\mathrm{Airy}}
\left(\frac{r_k}{\kappa},\frac{\kappa}{2}s_k;
\frac{r_l}{\kappa},\frac{\kappa}{2}s_l\right)}_{1\leq k,l\leq m}.
\end{equation}
Theorem~\ref{thmEdge} thus follows from (\ref{eqEdge.9b}).
\end{proof}
We turn to the proof of Proposition~\ref{propEdge}. As bounded set we fix throughout a centered box $\mathcal{B} \subset \R^d$, where the dimension $d$ depends on the context.

\begin{proof}[\sc{Proof of Proposition~\ref{propEdge}:}]
Let us first consider $s_2\geq s_1$. By definition of $K_T(r_2,s_2;r_1,s_1)$,
(\ref{eqLine.22}), (\ref{eqLine.23}), and (\ref{eqLine.28}), we have
\begin{eqnarray}\label{eqEdge.13}
& & \hspace{-40pt}K_T(r_2,s_2;r_1,s_1)= \frac{e^{-g(r_1,s_1)}}{e^{-g(r_2,s_2)}} T^{1/3} \times \nonumber \\
& \times&
\sum_{l\leq 0}\frac{1}{2\pi}\int_{-\pi}^\pi\dk{k} e^{-i k j(r_1,s_1)}e^{i k l}
e^{\sum_{n\geq 1}(\mu_n e^{i k n}-\nu_n e^{-i k n})}e^{\sum_{n\geq 1}\phi_n^1
e^{-i k n}} \\
& \times& \frac{1}{2\pi}\int_{-\pi}^\pi\dk{k'} e^{i k' j(r_2,s_2)}e^{-i k' l}
e^{-\sum_{n\geq 1}(\mu_n e^{i k' n}-\nu_n e^{-i k' n})}e^{-\sum_{n\geq 1}\phi_n^2
e^{-i k' n}}, \nonumber
\end{eqnarray}
where $\mu_n=q^{n/2}/n(1-q^n)$, $\nu_n=\mu_n q^{n \tau_0 T}$, and
$\phi_n^i=\nu_n(1-q^{n s_i T^{2/3}})$. As in Section~\ref{bulk} we regard the integrals in (\ref{eqEdge.13}) as complex line integrals and use the notation  explained below (\ref{eqBulk.11}).

The integrands in (\ref{eqEdge.13}) are holomorphic away from
$\{k\in \C \, | \, \Re(k)=0 , |\Im(k-i\tau_0/2)|\geq\tau_0/2\}$ and the straight path from $-\pi$ to $\pi$ can be deformed provided no singularities are touched. 
In our choice the deformed path has three straight lines,
the first one from $-\pi$ to $-\pi+i\beta_i(T)$, the second one from
$-\pi+i\beta_i(T)$ to $\pi+i\beta_i(T)$, and the last one from
$\pi+i\beta_i(T)$ to $\pi$ with $\beta_i \in (0,\tau_0)$, see Figure~\ref{FigEdge}.
To be precise, the path along the real line touches at $k=0$ the starting point of a branch cut of the term in the exponential, but still the integral remains unchanged by the above deformation.
\begin{figure}[t!]
\begin{center}
\psfrag{Re}[][][1]{$\Re$}
\psfrag{Im}[][][1]{$\Im$}
\psfrag{0}[][][1]{$0$}
\psfrag{t}[][][1]{$\tau_0$}
\psfrag{mp}[][][1]{$-\pi$}
\psfrag{pp}[][][1]{$\pi$}
\includegraphics[height=5cm]{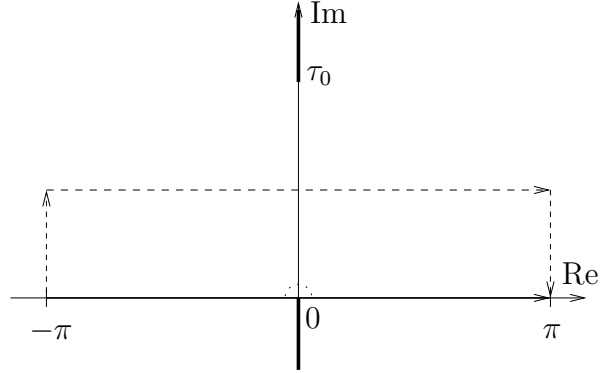}
\caption{\it Deformation of the integration path. The original integral, from $-\pi$ to $\pi$, is deformed along the integral on the dashed path.}
\label{FigEdge}
\end{center}
\end{figure}
Since the integrands are $2\pi$-periodic along the real axis, the first and the last integrals cancels exactly. $\beta_1(T)$ is determined such that the terms
in the exponential are purely imaginary. We obtain
\begin{equation}
\beta_i(T)=-\frac{1}{2}\left(s_i T^{2/3}\ln(1-1/T)+\tau_0 T \ln(1-1/T)\right), \quad i=1,2.
\end{equation}
We also define $l=L T^{1/3}$. Then the summation goes over
$L \in T^{-1/3}(\Z_{-}\cup\{0\})$ and
\begin{equation}
K_T(r_2,s_2;r_1,s_1)= \frac{e^{-g(r_1,s_1)}}{e^{-g(r_2,s_2)}} \frac{T^{1/3}}
{4\pi^2} 
\sum_{L \in T^{-1/3}(\Z_{-}\cup\{0\})}\hspace{-0.8cm}
\wt{J}_1(L) \wt{J}_2(L) e^{\delta_1-\delta_2},
\end{equation}
where
\begin{equation}
\delta_i=j(r_i,s_i) \beta_i(T)-\beta_i(T) L T^{1/3}
\end{equation} and
\begin{eqnarray}
\wt{J}_1(L)&=&\int_{-\pi}^{\pi}{e^{-i k j(r_1,s_1)}e^{i k L T^{1/3}} \exp\bigg[2 i \sum_{n\geq 1}{\mu_n \sin(k n)e^{-\beta_1(T) n}}\bigg]\dk{k}},\\
\wt{J}_2(L)&=&\int_{-\pi}^{\pi}{e^{i k' j(r_2,s_2)}e^{-i k' L T^{1/3}} \exp\bigg[-2 i \sum_{n\geq 1}{\mu_n \sin(k' n)e^{-\beta_2(T) n}}\bigg]\dk{k'}}. \nonumber
\end{eqnarray}
Finally defining $J_i(L)=T^{1/3} \wt{J}_i(L)$, we have
\begin{equation}\label{eqEdge.18}
K_T(r_2,s_2;r_1,s_1) = \hspace{-0.6cm}
\sum_{L \in T^{-1/3}(\Z_{-}\cup\{0\})}\hspace{-0.6cm}
(4\pi^2 T^{1/3})^{-1}
e^{\frac{1}{2}L (s_2-s_1)\left(1+\Or{T^{-1}}\right)} J_1(L) J_2(L).
\end{equation}
For the case $s_2<s_1$ the result is
\begin{equation}
K_T(r_2,s_2;r_1,s_1) = - \hspace{-0.6cm}\sum_{L \in T^{-1/3}(\Z_{+}\setminus\{0\})}
\hspace{-0.6cm}(4\pi^2 T^{1/3})^{-1}
e^{\frac{1}{2}L (s_2-s_1)\left(1+\Or{T^{-1}}\right)} J_1(L) J_2(L).
\end{equation}

Now we proceed as follows. First we prove that, as $T\to\infty$,
$J_i(L)\to \frac{2\pi}{\kappa}\Ai{\frac{r_i-L}{\kappa}}$ for $L\in\mathcal{B}$, by using the steepest descend curve for the term which is exponentially small in $T$. Secondly we consider separately $s_2<s_1$ and $s_2\geq s_1$. In the
latter case, for large $L$, we need the steepest descend curve for the whole integrand. The same strategy has been used in~\cite{GTW}.
In the case $s_2<s_1$, for large $L$, the steepest descend curve does not
exist anymore. On the other hand the term $e^{-L(s_1-s_2)}$ serves as a convergence factor and we only need to find bounds for the $J_i(L)$.\\

\noindent\textbf{Convergence for $L$ in a bounded set.}\\
Let $L\in\mathcal{B}$. The integral $J_1(L)$ is written as
\begin{equation}\label{eqEdge.19}
J_1(L)=T^{1/3}\int_{-\pi}^{\pi}{e^{T\psi_{1,T}(k)}e^{i k L T^{1/3}}}\dk{k},
\end{equation}
where 
\begin{equation}
\psi_{1,T}(k)=-i k T^{-1} j(r_1,s_1)+
2i\sum_{n\geq 1}{\frac{\mu_n}{T} e^{-\beta_1 n}\sin(kn)}.
\end{equation}
We make a saddle point approximation by using a curve which, for small $k$,
is very close to the steepest descend curve for $\psi(k)$, where
\begin{equation}
\psi(k)=\lim_{T\to\infty}{(i k L T^{1/3}+\psi_{1,T}(k))/T}
\end{equation}
and the convergence is uniform for $(s_1,r_1,L)\in\mathcal{B}$.
For the limit we obtain
\begin{equation}
\psi(k)=\psi_0(k)+2i k \ln\left(1-e^{-\tau_0/2}\right),
\end{equation}
where
\begin{equation}
\psi_0(k)=\sum_{n\geq 1}{2 i \sin(kn)\frac{e^{-n\tau_0/2}}{n^2}}.
\end{equation}
In particular $\psi(k)$ is holomorphic in 
$\C \setminus \{k=x+iy\in \C\, | \, x=0 ,\abs{y}\geq \tau_0/2\}$
and the whole integrand is $2\pi$-periodic along the real axis.

Instead of integrating along the straight path $-\pi\to\pi$ we integrate
along $C=\{k=x+iy,y=-\abs{x}/\sqrt{3}\}$, see Figure~\ref{FigEdge2}.
\begin{figure}[t!]
\begin{center}
\psfrag{Re}[][][1]{$\Re$}
\psfrag{Im}[][][1]{$\Im$}
\psfrag{0}[][][1]{$0$}
\psfrag{t}[][][1]{$\frac{\tau_0}{2}$}
\psfrag{tt}[][][1]{$-\frac{\tau_0}{2}$}
\psfrag{mp}[][][1]{$-\pi$}
\psfrag{pp}[][][1]{$\pi$}
\psfrag{C}[][][1]{$C$}
\psfrag{arg}[][][1]{$\pi/6$}
\includegraphics[height=5cm]{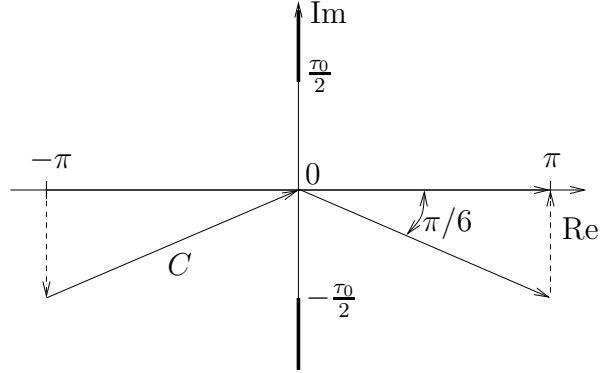}
\caption{\it Deformation of the integration path. The path from $-\pi$ to $\pi$ is deformed into $C$ plus the dashed ones.}
\label{FigEdge2}
\end{center}
\end{figure}
For $x$ small this path is almost at steepest descend.
The real part of $\psi(k)$ reaches its maximum at $k=0$.
To evaluate the errors for $x$ away from zero we prove that the real part of
$\psi(k)$ is strictly decreasing for $\abs{x}$ increasing. By symmetry we
consider only $x\in [0,\pi]$. A simple computation gives
\begin{equation}
\Dt{\psi(k)}{x}=-(i+1/\sqrt{3})\ln(Q)
\end{equation}
with
\begin{equation}
Q=\frac{(1-e^{i x + x/\sqrt{3}-\tau_0/2})(1-e^{-i x-x/\sqrt{3}-\tau_0/2})}{(1-e^{-\tau_0/2})^2}
\end{equation}
and
\begin{eqnarray}
\Re{Q}&=&\frac{\cosh(\tau_0/2)-\cosh(x/\sqrt{3})\cos(x)}{2\sinh^2(\tau_0/4)},\nonumber\\
\Im{Q}&=&-\frac{\sin(x)\sinh(x/\sqrt{3})}{2\sinh^2(\tau_0/4)}.
\end{eqnarray}
Using that $\cosh(x/\sqrt{3})\cos(x)\leq 1$ and is maximal at $x=0$, we
have $\Re{Q(x)}\geq\Re{Q(0)} =1$, the inequality being strict if $x\neq 0$.
Obviously $\Im{Q}\leq 0$. Therefore
\begin{equation}\label{eqEdge.25}
\Re\left(\Dt{\psi(k)}{x}\right)=-\frac{1}{2\sqrt{3}}\ln((\Re{Q})^2+(\Im{Q})^2)+
\arctan(\Im{Q}/\Re{Q})\leq 0
\end{equation}
for all $x \in [0,\pi]$ and for all $\tau_0\in (0,\infty)$. The inequality is
strict if $x\neq 0$. Since $\Dt{\Re(\psi(k))}{x}=\Re\left(\Dt{\psi(k)}{x}\right)$
and by (\ref{eqEdge.25}), $\Re{\psi(k)}$ is maximal at $k=0$, $\psi(0)=0$, and is strictly
decreasing for $\abs{x}$ increasing.

Let us fix $\e$, $0<\e\ll 1$, and let $C_\e$ be the part of $C$ with $x\in [-\e,\e]$.
Then the contribution at $J_1(L)$ coming from $C\setminus C_\e$ is exponentially
small in $T$.
\begin{lem}\label{lemEdge.3} For some $\delta>0$,
\begin{equation}
J_1(L)=\Or{e^{-\delta T}} +
T^{1/3}\int_{C_\e}{e^{\psi_{1,T}(k)T}e^{i k L T^{1/3}}}\dk{k}.
\end{equation}
\begin{proof}[\sc{Proof:}]
Let $\wt{C}_\e^+$ be the part of $C$ with $x\in [\e,\pi]$ and $\wt{C}_\e^-$
the one with $x \in [-\pi,-\e]$. For $x \geq \e$, $\Re\psi(k)\leq
\Re\psi(0)-2\delta <0$ for suitable $\delta=\delta(\e)>0$. In addition
\begin{equation}
\psi_{1,T}(k)T+i k L T^{1/3}=\psi(k) T+\Or{(L-r_1) T^{1/3}+s_1 T^{2/3}}.
\end{equation}
Then
\begin{equation}
\abs{\int_{\wt{C}_\e^+}e^{\psi_{1,T}(k)T}e^{i k L T^{1/3}}\dk{k}} \leq
e^{-\delta T}\int_{\e}^{\pi}{\frac{2}{\sqrt{3}}e^{(\Re\psi(k)-\delta)T}
e^{\Or{(L-r_1) T^{1/3}+s_1 T^{2/3}}}\dk{x}}.
\end{equation}
For $(L,s_1,r_1)\in\mathcal{B}$, the integral on the right side is uniformly 
bounded and therefore
\begin{equation}
\abs{\int_{\wt{C}_\e^+}e^{\psi_{1,T}(k)T}e^{i k L
T^{1/3}}\dk{k}}=\Or{e^{-\delta T}}.
\end{equation}
Similarly for the integral along $\wt{C}_\e^-$.
\end{proof}
\end{lem}

\begin{lem}\label{uniformly} Uniformly for $(L,r_1,s_1)\in\mathcal{B}$, one has
\begin{equation}\label{eqEdge.31}
J_1(L)=\Or{e^{-\delta T}}+
\Or{T^{-1/3}}+\frac{2\pi}{\kappa}\Ai{\frac{r_1-L}{\kappa}}
\end{equation}
for large $T$, with $\kappa=\sqrt[3]{2\alpha_3}$.
\begin{proof}[\sc{Proof:}]
By Lemma~\ref{lemEdge.3} we have to evaluate the contribution of the integral
along $C_\e$. For $k$ close to $0$ we have
\begin{eqnarray}\label{eqEdge.29}
\psi_{1,T}(k)T+i k L T^{1/3}&=&-\frac{2}{3}i\alpha_3 k^3 T -i k
T^{1/3}(r_1-L)\\ & &+\,\, \Or{s_1^2 k+s_1 k^3 T^{2/3}+k^5 T}.\nonumber
\end{eqnarray}
Let $C_\e^+$ be the part of $C$ with $x\in [0,\e]$ and $C_\e^-$ the one
with $x\in [-\e,0]$. Then $\int_{C_\e}{\cdots}=
\int_{C_\e^+}{\cdots}+\int_{C_\e^-}{\cdots}.$
We consider explicitly only one of the two integrals, the second being evaluated 
in the same fashion,
\begin{eqnarray}
& &\hspace{-40pt}T^{1/3}\int_{C_\e^+}{e^{\psi_{1,T}(k)T+i k L T^{1/3}}\dk{k}}
\nonumber \\
& =&
T^{1/3}\int_{C_\e^+}{e^{-\frac{2}{3} i \alpha_3 k^3 T}
e^{-i k T^{1/3}(r_1-L)}e^{\Or{s_1^2 k+s_1 k^3T^{2/3}+k^5T}}\dk{k}} \\
& =&T^{1/3}\int_{C_\e^+}{e^{-\frac{2}{3}i \alpha_3 k^3
T}e^{-i k T^{1/3}(r_1-L)}\dk{k}}+E_1(L). \nonumber
\end{eqnarray}
The error term is the integral along $C_\e^+$ with integrand
\begin{eqnarray}
& &\hspace{-20pt}T^{1/3}{e^{-\frac{2}{3}i \alpha_3 k^3 T}
e^{-i k T^{1/3}(r_1-L)}(e^{\Or{s_1^2 k+s_1 k^3T^{2/3}+k^5T}}-1)}\\
&=&\!\!\!T^{1/3}{e^{-\frac{2}{3}i \alpha_3 k^3 T}
e^{-i k T^{1/3}(r_1-L)}e^{\Or{s_1^2 k+s_1 k^3T^{2/3}+k^5T}}
\Or{s_1^2 k+s_1 k^3T^{2/3}+k^5T}}\nonumber.
\end{eqnarray}
The term in the exponential is
$-\frac{2}{3}i \alpha_3 k^3 T (1+\chi_1)-i k T^{1/3}(r_1-L)(1+\chi_2)$, where
$\chi_1$ and $\chi_2$ can be made arbitrarily small by taking $\e$ small enough
($s_1$ is bounded).
With the change of variable $z=kT^{1/3}$ we obtain
\begin{equation}
E_1(L)=\frac{1}{T^{1/3}}\int_{T^{1/3}C_\e^+}
{\hspace{-0.8cm}e^{-i\frac{2}{3}\alpha_3(1+\chi_1)z^3-i(1+\chi_2)(r_1-L)z}
\Or{s_1^2 z+s_1 z^3+z^5 T^{-1/3}}\dk{z}}.
\end{equation}
Remark that at the boundary of the integration, the real part of the
integrand behaves as $e^{-\frac{2}{3} \alpha_3 \e^3 T}$. This integral is
uniformly bounded in $T$ for $(L,r_1,s_1)\in\mathcal{B}$. The same holds
for the integral on $C_\e^-$. Consequently $E_1(L)=\Or{T^{-1/3}}$.

Next we extend the integration from $C_\e$ to $-\pi T^{1/3}(1,\cos(\pi/6))$
and $\pi T^{1/3}(1,-\cos(\pi/6))$, obtaining the path $D_1$. In this way
we add an error of $\Or{e^{-\delta'(\e)T}}$ with $\delta'(\e)\sim \e^3$. Similarly
we can complete the path up to $x=\pm N \pi T^{1/3}$, $y=-N \pi T^{1/3}/\sqrt{3}$ by straight lines. The integral is equal to the integration from $-N \pi T^{1/3}$ to $N\pi T^{1/3}$, since the function is $2\pi T^{1/3}$ periodic in the real
direction and the error added by completing the integral is exponentially small
in $T$, for all $N$. Therefore we may take the limit $N\to\infty$.

Finally we obtain (\ref{eqEdge.31}), since 
\begin{equation}
\int_{-\infty}^{\infty}{e^{-i\frac{2}{3}\alpha_3
z^3 - i z (r_1-L)}\dk{z}}=\frac{2\pi}{\kappa}\Ai{\frac{r_1-L}{\kappa}}
\end{equation}
with $\kappa=\sqrt[3]{2\alpha_3}$.
\end{proof}
\end{lem}

\noindent\textbf{Convergence of $K_T(r_2,s_2;r_1,s_1)$ with $s_2<s_1$.}
\begin{lem}
Uniformly for $(s_i,r_i)\in\mathcal{B}$, $i=1,2$,
\begin{equation}\label{eqEdge.34}
\lim_{T\to\infty}{K_T(r_2,s_2;r_1,s_1)}=
-\int_{0}^{\infty}{e^{\frac{1}{2} L (s_2-s_1)}\Ai{\frac{r_1-L}{\kappa}}
\Ai{\frac{r_2-L}{\kappa}}\frac{\dk{L}}{\kappa^2}}
\end{equation}
with $\kappa=\sqrt[3]{2\alpha_3}$.

\begin{proof}[\sc{Proof:}]
Since $(r_1,r_2)\in\mathcal{B}$, let us set $L_0$ such that
$L_0\leq 2(\abs{r_1}+\abs{r_2}+1)$ for all $r_1,r_2$.
$K_T$ can be transformed into an integral adding an error
$\Or{T^{-1/3}}$. Let us fix an $\e$ with $0<\e\ll 1$. Then 
\begin{eqnarray}
-K_T(r_2,s_2;r_1,s_1)&=&\int_{0}^{L_0}J_1(L) J_2(L) \frac{e^{-L X}}{4\pi^2}
\dk{L}+
\int_{L_0}^{\e T^{2/3}} \hspace{-12pt}J_1(L) J_2(L) \frac{e^{-L X}}{4\pi^2} \dk{L} \nonumber\\
& &+\,\,\int_{\e T^{2/3}}^\infty J_1(L) J_2(L) \frac{e^{-L X}}{4\pi^2} \dk{L}+\Or{T^{-1/3}}
\end{eqnarray}
with $X=\frac{1}{2}(s_1-s_2) (1+\Or{1/T})>0$.
Since $\abs{J_i(L)}\leq T^{1/3}$, $i=1,2$, the third term
is bounded by $T^{2/3} e^{-\e T^{2/3} X}/X \to 0$ as $T\to\infty$.
By Lemma~\ref{uniformly} the first term converges, uniformly for $(u_i,s_i)\in\mathcal{B}$, to
\begin{equation}
\int_{0}^{L_0}\Ai{\frac{r_1-L}{\kappa}}\Ai{\frac{r_2-L}{\kappa}}
e^{L \frac{s_2-s_1}{2}}\frac{1}{\kappa^2}\dk{L}
\end{equation} as $T\to\infty$.

We consider the second term. We have already established the pointwise convergence of $J_i(L)$ to $\frac{2\pi}{\kappa}\Ai{\frac{r_i-L}{\kappa}}$. If we obtain that for large $T$, $\abs{J_i(L)}\leq G$ with a constant $G$ independent of $r_i,s_i$ and $L \in [L_0,\e T^{2/3}]$, then by dominated convergence
\begin{equation}
\lim_{T\to\infty}\int_{L_0}^{\e T^{2/3}}\hspace{-0.6cm} J_1(L) J_2(L) e^{-L X}\dk{L}=
\int_{L_0}^{\infty} \hspace{-0.2cm}\Ai{\frac{r_1-L}{\kappa}} \Ai{\frac{r_2-L}{\kappa}}
\frac{e^{\frac{1}{2}L (s_2-s_1)}}{\kappa^2}\dk{L}
\end{equation}
uniformly for $(r_i,s_i)\in\mathcal{B}$.
This property is proven in the following lemma.
\end{proof}
\end{lem}

\begin{lem} For $L \in [L_0,\e T^{2/3}]$,
$\abs{J_i(L)}\leq G$ with the constant $G$ independent of $s_i,r_i,$ and $L$,
provided $0<\e\ll 1$ and $T$ large enough.
\begin{proof}[\sc{Proof:}]
The exponential terms in (\ref{eqEdge.19}) are purely imaginary for $k$ real. Let
us set
\begin{equation}
\psi_1^I(k)=\frac{1}{i}(i k L T^{-2/3}+\psi_{1,T}(k)),
\end{equation}
then
\begin{equation}\label{eqEdge.40}
J_1(L)=T^{1/3}\int_{-\pi}^{\pi}e^{i \psi_1^I(k) T}\dk{k}.
\end{equation}
In particular for $k$ close to $0$,
\begin{equation}\label{eqEdge.45}
\psi_1^I(k)=-\frac{2}{3}\alpha_3 k^3\left(1+\Or{k^2\!+\!s_1 T^{-1/3}}\right)
-k(r_1-L)T^{-2/3}\left(1+\Or{s_1^2 T^{-1/3}}\right).
\end{equation}
Since $(r_1-L)T^{-2/3}\sim \Or{\e}$ at most, we set $\wt{L}=(L-r_1)T^{-2/3}$.
$\psi_1^I(k)$ has two local extrema at $\pm k(\wt{L})$ with
\begin{equation}
k(\wt{L})=\sqrt{\wt{L}}c_0\left(1+ \Or{\wt{L}+s_1^2 T^{-1/3}}\right)
\end{equation}
and $c_0=(2 \alpha_3)^{-1/2}$. Moreover for $\abs{k}\geq 2 k(\wt{L})$, $\psi_1^I(k)$ is strictly decreasing.
$J_1(L)=\int_{-\pi}^\pi \cdots = \sum_{i=1}^4 \int_{I_i} \cdots$ where
$I_1=[-\pi,-2 c_0 \sqrt{\wt{L}}]$, $I_2=[-2 c_0 \sqrt{\wt{L}},0]$,
$I_3=[0,2 c_0 \sqrt{\wt{L}}]$, and $I_4=[2 c_0 \sqrt{\wt{L}},\pi]$. The integrals
along $I_1$ and $I_4$ are evaluated similarly and so are the integrals along $I_2$
and $I_3$. We present in detail only the integration along $I_3$ and $I_4$.
Let $\gamma=\sqrt{\wt{L}}$. Then
\begin{equation}
\int_{I_4}\cdots =T^{1/3}\int_{2c_0 \gamma}^\pi e^{i \psi_1^I(k) T} \dk{k} =
T^{1/3} \int_{u(2 c_0 \gamma)}^{u(\pi)} f(u) e^{i u T} \dk{u}
\end{equation}
where $u=\psi_1^I(k)$ and $f(u)=\Dt{k(u)}{u}$. Integrating by parts we obtain
\begin{equation}
T^{-1/3} \int_{I_4}\cdots =f(u)\frac{e^{i u T}}{iT}\bigg\arrowvert_{u(2 c_0 \gamma)}^{u(\pi)}-\int_{u(2 c_0 \gamma)}^{u(\pi)} \Dt{f(u)}{u} \frac{e^{i u T}}{iT} \dk{u}.
\end{equation}
For $k\in I_4$ with $\abs{k}\leq \e$ follows from (\ref{eqEdge.45})
that $\Dt{u}{k}<0$ and $\Dtt{u}{k}\geq 0$. For $k>\e$,
\begin{equation}\label{eqEdge.49}
\Dt{u}{k} = \wt{L}-\ln\left(\frac{1+e^{-\tau_0}-2 e^{-\tau_0/2}\cos(k)}
{1+e^{-\tau_0}-2 e^{-\tau_0/2}}\right)+\Or{s_1 T^{-1/3}}.
\end{equation}
Then for $k\in I_4$ with $k>\e$, $\Dt{u}{k}<0$ and $\Dtt{u}{k}\geq 0$.
Therefore $\Dt{f(u)}{u}=-(\Dt{u}{k})^{-3}\Dtt{u}{k}$ where $\Dt{u}{k} < 0$ and 
$\Dtt{u}{k}\geq 0$ for every point in $I_4$.
Thus $\Dt{f(u)}{u}$ does not change sign along $I_4$, and 
\begin{equation}
\abs{\int_{I_4}\cdots}\leq \frac{2}{T^{2/3}}\left(\abs{f(u(\pi))}+\abs{f(u(2 c_0 \gamma))}\right).
\end{equation}
Using (\ref{eqEdge.49}), for $T$ sufficiently large, 
\begin{eqnarray}
\abs{f(u(\pi))}&=&\abs{2\ln(1-e^{-\tau_0/2})-2\ln(1+e^{-\tau_0/2})+\gamma^2+
\Or{s_1 T^{-1/3}}}^{-1}
\nonumber \\
&\leq& \abs{\ln(1-e^{-\tau_0/2})-\ln(1+e^{-\tau_0/2})}^{-1}=G_1,
\end{eqnarray}
provided $\e$ small enough (which implies $\gamma$ sufficiently small). The second term is bounded by
\begin{equation}
\abs{f(u(2 c_0 \gamma))} = \abs{\frac{1+\Or{\gamma^2+s_1 T^{-1/3}}}{-\gamma^2}}
\leq 2/\gamma^2
\end{equation} for $\e$ small and $s_1\in\mathcal{B}$. 
Therefore we have, uniformly in $(r_1,s_1)\in\mathcal{B}$,
\begin{equation}
\abs{\int_{I_4}\cdots}=\abs{\int_{I_1}\cdots}\leq \frac{2 G_1}{T^{2/3}}+
\frac{2}{(L-r_1)}\leq \frac{2 G_1}{T^{2/3}}+ \frac{2}{(L_0-r_1)}.
\end{equation}

Next we estimate $\abs{\int_{I_3}\cdots}$. 
\begin{equation}
\int_{I_3}\cdots=T^{1/3}\int_0^{2 c_0\gamma}{e^{i \psi_1^I(k) T} \dk{k}}=
T^{1/3}\int_{-c_1\gamma}^{c_2\gamma}{e^{i \wt{\psi}(k) T} \dk{k}}
\end{equation}
where $\wt{\psi}(k)=\psi_1^I(k-k(\wt{L}))$, $c_1=c_0(1+\Or{\gamma})$ and
$c_2=c_0(1+\Or{\gamma})$. Let us define the paths $\xi_0=\{k=x, x:-c_1 \gamma \to
c_2 \gamma\}$, $\xi_1=\{k= -c_1 \gamma e^{-i \phi}, \phi:0 \to \pi/4\}$,
$\xi_2=\{k=e^{-i \pi/4} x, x:-c_1 \gamma \to c_2 \gamma\}$, $\xi_3=\{k= c_2
\gamma e^{i \phi}, \phi:\pi/4 \to 0\}$. Then $\int_{I_3}\cdots=\int_{\xi_0}\cdots
=\sum_{i=1}^3\int_{\xi_i} \cdots$. The integrals along $\xi_1$ and $\xi_3$ are
estimated in the same way.
\begin{equation}
T^{1/3}\int_{\xi_1} e^{i \wt{\psi}(k) T} \dk{k}=
T^{1/3}\int_0^{\pi/4} e^{i\phi} e^{i \wt{\psi}(k(\phi))T} i c_0
\gamma (1+\Or{\gamma}) \dk{\phi}
\end{equation}
and therefore
\begin{equation}
\abs{T^{1/3}\int_{\xi_1} e^{i \wt{\psi}(k) T} \dk{k}} \leq
2 T^{1/3} \gamma c_0 \int_{0}^{\pi/4} e^{-T \Im(\wt{\psi}(k(\phi)))}\dk{\phi}.
\end{equation}
Since $\wt{\psi}(k(\phi))=\wt{\psi}(0)+\frac{1}{2}\wt{\psi}''(0) k(\phi)^2
(1+\delta_1(\phi))$ with $\delta_1(\phi) \to 0$ as $\e \to 0$ and
$k(\phi)^2=c_0^2\gamma^2(1+\Or{\gamma}) e^{-2i \phi}$, one has
\begin{equation}
\Im \wt{\psi}(k(\phi)) = -\frac{1}{2}\wt{\psi}''(0) (k(\phi))^2
(1+\delta_2(\phi)) c_0^2 \gamma^2 (1+\Or{\gamma}) \sin(2\phi)
\end{equation}
with $\delta_2(\phi)\to 0$ as $\e\to 0$.
Moreover, for $\e$ small enough, $\sin(2\phi)(1+\delta_2(\phi))(1+\Or{\gamma}) \geq \phi$.
From this it follows
\begin{eqnarray}
\abs{T^{1/3}\int_{\xi_1} e^{i\wt{\psi}(k) T} \dk{k}} &\leq&
2 T^{1/3} c_0 \gamma \int_0^{\pi/4}{e^{T c_0^2 \gamma^2 \wt{\psi}''(0)
\phi/2}}\dk{\phi} \\
&\leq& 2 T^{1/3} c_0 \gamma \int_0^\infty{e^{T c_0^2 \gamma^2 \wt{\psi}''(0)
\phi/2}}\dk{\phi} = \frac{4 T^{1/3} c_0 \gamma}{T c_0^2 \gamma^2 \abs{\wt{\psi}''(0)}}.\nonumber
\end{eqnarray}
We compute
$\wt{\psi}''(0)=-2\gamma c_0^{-1}\left(1+\Or{\gamma^2+s_1 T^{-1/3}}\right)$.
Therefore for $s_1\in\mathcal{B}$ and $T$ large enough,
\begin{equation}
\abs{T^{1/3}\int_{\xi_1} e^{i \wt{\psi}(k) T} \dk{k}}
\leq  \frac{4}{(L-r_1)}\leq \frac{4}{(L_0-r_1)}.
\end{equation}

Next we need to evaluate the integral along $\xi_2$,
\begin{equation}
T^{1/3}\int_{\xi_2} e^{i\wt{\psi}(k) T} \dk{k} =
T^{1/3} e^{-i \pi/4} e^{i T \wt{\psi}(0)}\int_{-c_1 \gamma}^{c_2 \gamma}
e^{T \wt{\psi}''(0)x^2/2(1+\Or{x})}\dk{x}.
\end{equation}
Then for sufficiently small $\delta$,
\begin{eqnarray}
& &\hspace{-60pt}\abs{T^{1/3}\int_{\xi_2} e^{i\wt{\psi}(k) T} \dk{k}} \leq
T^{1/3} \int_{-c_1\gamma}^{c_2\gamma}e^{T \wt{\psi}''(0)x^2
(1+\delta)/2}\dk{x} \nonumber \\
&\leq& T^{1/3}\int_{-\infty}^\infty e^{T\wt{\psi}''(0)x^2
3/4}\dk{x} \leq T^{1/3} \int_{-\infty}^\infty e^{-x^2 T \gamma/c_0}\dk{x} \\
&=&\frac{\sqrt{\pi}\sqrt{c_0}}{\sqrt[4]{L-r_1}}
=\frac{\sqrt{\pi}\sqrt{c_0}}{\sqrt[4]{L_0-r_1}}. \nonumber
\end{eqnarray}
Thus we have, uniformly for $s_1\in\mathcal{B}$ and $T$ large enough,
\begin{equation}
\abs{\int_{I_3}\cdots}\leq \frac{8}{(L_0-r_1)}+
\frac{\sqrt{\pi}\sqrt{c_0}}{\sqrt[4]{L_0-r_1}}.
\end{equation}
Therefore $J_i(L)$ is bounded by
\begin{equation}
\abs{J_i(L)}\leq \frac{4 G_1}{T^{2/3}} +\frac{20}{(L_0-r_i)}+
\frac{2\sqrt{\pi}}{\sqrt[4]{2 \alpha_3 (L_0-r_i)}}.
\end{equation}
Since $L_0-r_i\geq 2$ and $X>0$, it follows 
that $J_1(L) J_2(L)e^{-L X}$ is bounded by an integrable function on
$[L_0,\e T^{2/3}]$ for $0<\e\ll 1$ and $T$ large enough.
\end{proof}
\end{lem}

\noindent \textbf{Convergence of $K_T(r_2,s_2;r_1,s_1)$ with $s_2\geq s_1$.}
\begin{lem}
Uniformly for $(s_i,r_i)\in\mathcal{B}$, $i=1,2$,
\begin{equation}
\lim_{T\to\infty}{K_T(r_2,s_2;r_1,s_1)}=
\int_{-\infty}^{0}{e^{\frac{1}{2}L(s_2-s_1)}\Ai{\frac{r_1-L}{\kappa}}\Ai{\frac{r_2-L}{\kappa}}\frac{\dk{L}}{\kappa^2}}
\end{equation}
with $\kappa=\sqrt[3]{2\alpha_3}$.
\begin{proof}[\sc{Proof:}]
Let us set $L_0$ such that $L_0 \geq 2 (\abs{r_1}+\abs{r_2}+1)$ for all
$(r_1,r_2)\in\mathcal{B}$. Then the sum in $K_T$ can be approximated by an integral
at the expense of an error $\Or{T^{-1/3}}$. Let us fix $\e$, $0<\e\ll 1$. Then
\begin{eqnarray}\label{eqEdge.60b}
K_T(r_2,s_2;r_1,s_1)&=&\int_{-L_0}^{0}J_1(L) J_2(L) \frac{e^{L X}}{4\pi^2} \dk{L}+
\int_{-\e T^{2/3}}^{-L_0} J_1(L) J_2(L) \frac{e^{L X}}{4\pi^2} \dk{L} \nonumber\\
& & +\,\,\int_{-\infty}^{-\e T^{2/3}} J_1(L) J_2(L) \frac{e^{L X}}{4\pi^2} \dk{L}+\Or{T^{-1/3}},
\end{eqnarray}
with $X=\frac{1}{2}(s_2-s_1)(1+\Or{1/T})\geq 0$.
The convergence of the first term has already been proved. Let us set
$\wt{L}=-(L-r_1) T^{-2/3}$. In the remainder of the proof we set
\begin{equation}
\psi(k)=\frac{1}{i}\psi_{1,T}(k)-k \wt{L}.
\end{equation}

First consider $\wt{L}\geq \e$. 
\begin{equation}
J_1(L)=T^{1/3}\int_{-\pi}^{\pi}e^{\psi(k)T}\dk{k}.
\end{equation}
With the change of variable $u=\psi(k)$, $f(u)=\Dt{k(u)}{u}$, and integration by parts, we have
\begin{equation}
\abs{J_1(L)} \leq T^{1/3}\frac{2\psi(\pi)}{T}
\max_{k\in [\psi(-\pi),\psi(\pi)]}\abs{\Dt{f(u)}{u}}.
\end{equation}
To compute $\abs{\Dt{f(u)}{u}}$ we use $\Dt{f(u)}{u}=-(\Dt{u}{k})^{-3}\Dtt{u}{k}$.
$\abs{\Dt{u}{k}}$ is (\ref{eqEdge.49}) with $\wt{L}$ replaced by $-\wt{L}$.
It is easy then to see that uniformly for $s_1\in\mathcal{B}$,
$\max_{k\in [\psi(-\pi),\psi(\pi)]}\abs{\Dt{f(u)}{u}} \leq G_1
\wt{L}^{-1}$ for $G_1=2/(\sinh(\tau_0/2) \e)^2<\infty$. Then for a suitable constant $G_2<\infty$,
\begin{equation}\label{eqEdge.69b}
\abs{J_1(L)}\leq G_2 (r_1-L)^{-1}.
\end{equation}
The same holds for $J_2$, therefore the third term in (\ref{eqEdge.60b}) is
bounded by
\begin{equation}
\int_{-\infty}^{-\e T^{2/3}} \frac{G_2^2}{(L+\abs{r_1}+\abs{r_2})^{2}}\dk{L},
\end{equation}
which is convergent for $T$ finite and vanishes for $T\to\infty$.

Finally we consider $0<\wt{L} \leq \e$. Let us set
$\beta=\sqrt{2(\cosh(\tau_0/2)-1)}$. We integrate over
$C=\left\{k=x+iy(x),y(x)=-\sqrt{y(0)^2+x^2/3}\right\}$, with $i y(0)$ the stationary point of $\psi(\cdot,\wt{L})$, see Figure~\ref{FigEdge3}. $y(0)=-\beta\sqrt{\wt{L}}+\Or{\wt{L}^{3/2}}$ and $C$ is almost the steepest descend curve for $x$ small.
\begin{figure}[t!]
\begin{center}
\psfrag{Re}[][][1]{$\Re$}
\psfrag{Im}[][][1]{$\Im$}
\psfrag{0}[][][1]{$0$}
\psfrag{t}[][][1]{$\frac{\tau_0}{2}$}
\psfrag{tt}[][][1]{$-\frac{\tau_0}{2}$}
\psfrag{mp}[][][1]{$-\pi$}
\psfrag{pp}[][][1]{$\pi$}
\psfrag{C}[][][1]{$C$}
\psfrag{y0}[][][1]{$y_0$}
\includegraphics[height=5cm]{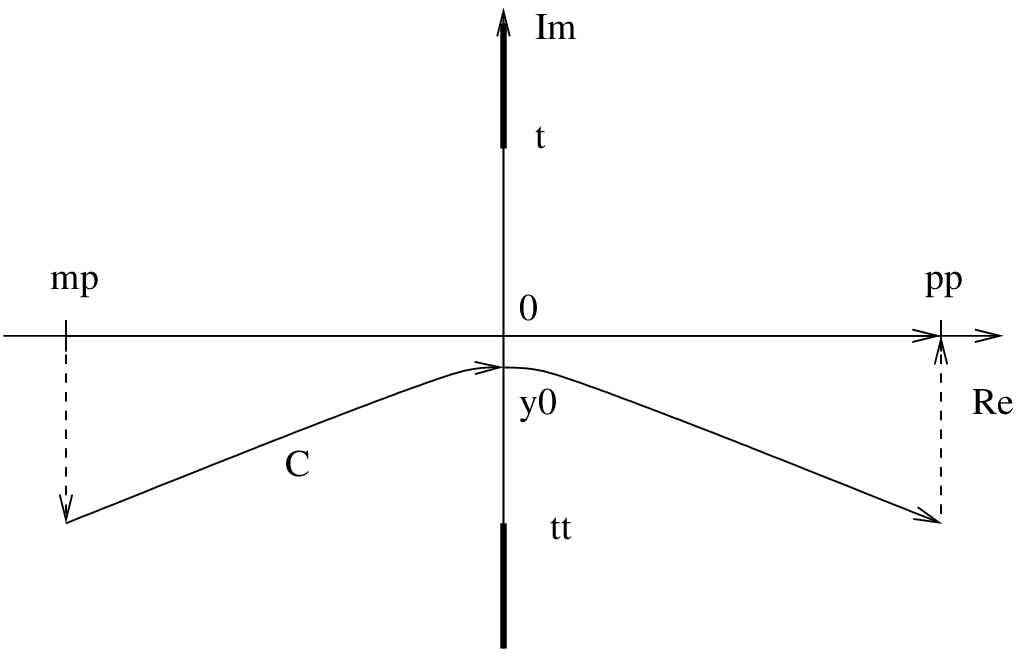}
\caption{\it Deformation of the integration path. The path from $-\pi$ to $\pi$ is deformed into $C$ plus the dashed ones.}
\label{FigEdge3}
\end{center}
\end{figure}
This path has the property that the real part of $\psi(k)$ is strictly decreasing as $\abs{x}$ increases and
\begin{equation}\label{eqEdge.65}
\psi(i y(0))=-\frac{2}{3}\beta \wt{L}^{3/2}\left(1+\Or{s_1^2
T^{-1/3}+\wt{L}}\right).
\end{equation}
We divide the integral in the part with $\abs{x}\leq \e$ and the remainder,
\begin{equation}
J_1(L)=T^{1/3}\int_{-\pi}^{\pi}{e^{\psi(k) T}\dk{k}}
= T^{1/3}\int_{C}{e^{\psi(k) T}\dk{k}}
= T^{1/3}\int_{C_\e}{e^{\psi(k) T}\dk{k}}+E_2(L),
\end{equation}
where 
\begin{equation}\label{eqEdge.60}
E_2(L)=T^{1/3}\int_{C\setminus C_\e}{e^{\psi(k) T}\dk{k}}
=\Or{e^{-\delta T}e^{\psi(x=0) T}} = 
\Or{e^{-\delta T}e^{-\frac{2}{3} \beta (r_1-L)^{3/2}}}.
\end{equation}
We then need to integrate only close to $x=0$. We first establish some
properties of $\psi(k)$ for $x=0$.

\begin{lem}
\begin{equation}\begin{array}{llcl} i)& \psi(iy(0),\wt{L})&=&-\frac{2}{3} \beta\wt{L}^{3/2}
+\Or{\wt{L}^{5/2}+\wt{L}^{3/2}s_1^2 T^{-1/3}},\\
ii) & \Dt{\psi(k,\wt{L})}{k}\arrowvert_{k=i y(0)}&=&0,\\
iii)& \Dtt{\psi(k,\wt{L})}{k}\arrowvert_{k=i y(0)}&=&-\frac{2}{\beta}
\sqrt{\wt{L}}+\Or{\wt{L}+\sqrt{\wt{L}}s_1 T^{-1/3}},\\
iv) & \Dttt{\psi(k,\wt{L})}{k}\arrowvert_{k=i y(0)}&=&-\frac{2}{\beta^2}i +
\Or{\wt{L}+s_1 T^{-1/3}}.
\end{array}
\end{equation}
\begin{proof}[\sc{Proof:}]
i) follows from Equation (\ref{eqEdge.65}) and ii) because $k=iy(0)$ is
a stationary point of $\psi(\cdot,\wt{L})$. iv) follows from (\ref{eqEdge.29}) because $2\alpha_3=1/\beta^2$.
Finally let $\lambda=\sqrt{\wt{L}}$. Then
\begin{equation}
\Dt{}{\lambda}\Dtt{\psi(k,\wt{L})}{k} =\Dttt{\psi(k,\wt{L})}{k} \Dt{k}{\lambda}
\end{equation}
and evaluating at $k=iy(0)$ and $\lambda=0$ we obtain iii).
\end{proof}
\end{lem}
With these properties 
\begin{eqnarray}
& &\hspace{-20pt}J_1(L)-E_2(L)=T^{1/3}\int_{C_\e}{e^{\psi(k) T}\dk{k}}\\
& &= e^{-\frac{2}{3} \beta \wt{L}^{3/2} T}e^{\Or{\wt{L}^{5/2} T +
\wt{L}^{3/2} s_1^2 T^{2/3}}}T^{1/3} \int_{C_\e}{\dk{k}e^{-\frac{1}{\beta}
\sqrt{\wt{L}} (k-i y(0))^2 T}e^{-\frac{i}{3\beta^2}(k-i y(0))^3 T}} 
\nonumber\\
& &\times e^{\Or{\wt{L}(k-i y(0))^2 T+\sqrt{\wt{L}} (k-i y(0))^2 s_1 T^{2/3}+
\wt{L}(k-i y(0))^3 T+(k-i y(0))^3 s_1 T^{2/3}+(k-i y(0))^4 T}}. \nonumber
\end{eqnarray}
Let $\gamma=\sqrt{r_1-L}$, then $\sqrt{\wt{L}}=\gamma T^{-1/3}$.
Let $k'=k-i y(0)$, then the integration is along $C'_\e=C_\e+i y(0)$.
\begin{eqnarray}\label{eqEdge.70}
& &\hspace{-20pt}J_1(L)-E_2(L)=e^{-\frac{2}{3} \beta \gamma^3}
e^{\Or{\gamma^5 T^{-2/3}+\gamma^3 s_1^2 T^{-1/3}}} T^{1/3}\int_{C'_\e}{\dk{k}
e^{-\frac{\gamma}{\beta} k^2 T^{2/3}}e^{-\frac{i}{3\beta^2} k^3 T}}\nonumber\\
& &\times \,
\exp[\Or{\gamma^2 k^2 T^{1/3}+\gamma s_1 k^2 T^{1/3}+\gamma^2 k^3 T^{1/3}+k^3
s_1 T^{2/3}+k^4 T}].
\end{eqnarray}
Since $\wt{L}$ can be made arbitrarily small, for $s_1\in\mathcal{B}$
the exponent of the term in the integral can be written as
\begin{equation}
-\frac{\gamma}{\beta} k^2 T^{2/3} (1+\chi_1) -\frac{i}{3\beta^2}k^3 T (1+\chi_2),
\end{equation}
where the $\chi_i$ can be made as small as desired by choosing $\e$
small enough. After the change of variable $k T^{1/3}=z$ the integral
becomes
\begin{equation}
\int_{C'_\e T^{1/3}}{\hspace{-0.6cm}\dk{z}e^{-\frac{\gamma}{\beta} z^2
(1+\chi_1)}e^{-\frac{i}{3\beta^2} z^3 (1+\chi_2)}}.
\end{equation}
The integration is taken along a contour, symmetric with respect to the imaginary
axis and such that for $\Re(z)\geq 0$, $\textrm{arg}(z)\in [-\pi/6,0]$.
This implies that the integral is uniformly bounded.

Replacing the term in front of the integral (\ref{eqEdge.70}) by one, the error can be estimated as 
\begin{equation}
e^{-\frac{2}{3} \beta \gamma^3}\left(e^{\Or{\gamma^5 T^{-2/3}+\gamma^3 s_1^2 T^{-1/3}}}-1\right),
\end{equation}
since the integral in (\ref{eqEdge.70}) is bounded. For $\wt{L}\leq \e$,
\begin{equation}
(r_1-L)^{5/2} T^{-2/3}+(r_1-L)^3 T^{-1/3} \leq
(r_1-L)^{3/2}\e + (r_1-L) \sqrt{\e}.
\end{equation}
As a consequence 
\begin{eqnarray}
e^{-\frac{2}{3} \beta \gamma^3}\left(e^{\Or{\gamma^5 T^{-2/3}+\gamma^3 s_1^2 T^{-1/3}}}-1\right)
&\leq& \Or{e^{-\frac{\beta}{2} \gamma^3}(\gamma^5 T^{-2/3}+\gamma^3 s_1^2 T^{-1/3})}
\nonumber\\
&\leq& \Or{T^{-1/3} e^{-\frac{\beta}{2}(r_1-L)^{3/2}}}. 
\end{eqnarray}
After this step we can also remove the error inside the integral (\ref{eqEdge.70}). As in the case of $L\in\mathcal{B}$, the removal of this error leads to an additional error of $T^{-1/3}$ with the prefactor
$e^{-\frac{2}{3}\beta (r_1-L)^{3/2}}$.
Consequently we have obtained
\begin{eqnarray}\label{eqEdge.83}
J_1(L)&=&e^{-\frac{2}{3}\beta (r_1-L)^{3/2}}
\int_{C'_\e T^{1/3}}{\hspace{-0.6cm}\dk{z}e^{-\frac{\gamma}{\beta}
z^2}e^{-\frac{i}{3\beta^2} z^3}}+\Or{T^{-1/3}e^{-\frac{2}{3}\beta 
(r_1-L)^{3/2}}} \nonumber\\
&+&\Or{T^{-1/3}e^{-\frac{1}{2}\beta (r_1-L)^{3/2}}}+
\Or{e^{-\delta T}e^{-\frac{2}{3}\beta (r_1-L)^{3/2}}}.
\end{eqnarray}
Next we change to the variable $z=w+i\beta\sqrt{r_1-L}$.
The integral becomes
\begin{equation}
\int_{C_\e'T^{1/3}+i\beta\gamma}\hspace{-1cm}e^{-\frac{i}{3\beta^2} w^3-i (r_1-L) w}\dk{w}.
\end{equation}

Finally completing the contour of the integration such that it goes to
infinity in the directions $\textrm{arg}(w)=\phi_{\pm}$ with
$\phi_+=-\pi/6$ and $\phi_-=-5\pi/6$ leads to an exponentially small
error. Using that $2\alpha_3=1/\beta^2$, the main term goes to 
$\frac{2\pi}{\kappa}\Ai{\frac{r_1-L}{\kappa}}$.
Since the errors are integrable in $L$ and go to zero as $T\to\infty$, we 
obtain, for $s_2\geq s_1$,
\begin{equation}\label{eqEdge.69}
\lim_{T\to\infty}{K_T(r_2,s_2;r_1,s_1)}=
\int_{-\infty}^{0}{e^{\frac{1}{2}L(s_2-s_1)}\Ai{\frac{r_1-L}{\kappa}}\Ai{\frac{r_2-L}{\kappa}}\frac{\dk{L}}{\kappa^2}}
\end{equation}
with $\kappa=\sqrt[3]{2\alpha_3}$.
\end{proof}
\end{lem}
With the change of variable $\lambda=L/\kappa$, (\ref{eqEdge.69}) is rewritten as
\begin{eqnarray}
\lim_{T\to\infty}{K_T(r_2,s_2;r_1,s_1)}&=&\kappa^{-1}
\int_{-\infty}^{0}{e^{\frac{1}{2}\lambda(s_2-s_1)\kappa}\Ai{\frac{r_1}{\kappa}-\lambda}
\Ai{\frac{r_2}{\kappa}-\lambda}\dk{\lambda}} \nonumber \\
&=& \kappa^{-1}K^{\mathrm{Airy}}
\left(\frac{r_2}{\kappa},\frac{\kappa}{2}s_2;
\frac{r_1}{\kappa},\frac{\kappa}{2}s_1\right).
\end{eqnarray}
\end{proof}

\section{The border step, Airy process}\label{border}
As explained in~\cite{PS}, the Airy field is a random field which is
concentrated on line ensembles $\{h_\ell(t),t\in \R, \ell \in \Z_+\}$ with the
properties\\[6pt]
i) $t\mapsto h_\ell(t)$ is continuous,\\[6pt]
ii) $h_\ell(t) < h_{\ell-1}(t)$ for all $t$.\\[6pt]
The first line, $h_0(t)$, of the Airy field is by definition the Airy process,
denoted by $A(t)$. $A(t)$ is almost surely continuous, stationary in $t$, and
invariant under time-reversal. Its single time distribution is given by the
Tracy-Widom distribution~\cite{TW}, known from the largest eigenvalue of GUE
random matrices. In particular, for fixed $t$,
\begin{eqnarray}
\Pb{A(t)>y} &\simeq& e^{-y^{3/2}4/3}\quad\textrm{for }y\to\infty,\nonumber\\
\Pb{A(t)<y} &\simeq& e^{-\abs{y}^3/12}\quad\textrm{for }y\to-\infty.
\end{eqnarray}
The Airy process is localized. On the other hand it has long range correlations as 
\begin{equation}
\ff{A(0)A(t)}-\ff{A(0)}^2\simeq \abs{t}^{-2}\quad\textrm{for }\abs{t}\to \infty.
\end{equation}

The convergence of $\eta_T^\mathrm{edge}$ to the Airy field, as stated in
Theorem~\ref{thmEdge}, implies that the border step statistics, properly scaled
as $A_T$, cf.\ (\ref{eqIntro.12}), converges to the Airy process.
\begin{thm}\label{thmBorder}
Let $A_T(s)$ be the border step rescaled as in (\ref{eqIntro.12})
and let $A(s)$ be the Airy process.
Then for any $m, s_i,a_i \in \R$, $i=1,\ldots,m$, the limit
\begin{equation}\label{eqBorder.1}
\lim_{T\to\infty}\PbT{\bigcap_{i=1}^{m}\{A_T(s_i)\leq a_i\}} =
\Pb{\bigcap_{i=1}^{m}\{A(s_i\kappa/2)\leq a_i/\kappa \}}
\end{equation}
holds.
\begin{proof}[\sc{Proof:}]
Let $f_i$ be the indicator function of $(a_i,\infty)$. Then (\ref{eqBorder.1})
corresponds to
\begin{equation}\label{eqBorder.2}
\lim_{T\to\infty}\PbT{\bigcap_{i=1}^{m}\{\eta_T^{\textrm{edge}}(f_i,s_i)=0\}}
=\Pb{\bigcap_{i=1}^{m}\{\eta^{\textrm{Airy}}(f_i/\kappa ,s_i \kappa/2)=0\}}.
\end{equation}
We choose $a$ large enough and split $f_i=f_i^a+g^a$ with $f_i^a$ the indicator
function of $(a_i,a]$ and $g^a$ the one of $(a,\infty)$.
Then
\begin{eqnarray}\label{eqBorder.5}
\bigg|\PbT{\bigcap_{i=1}^{m}\{\eta_T^{\textrm{edge}}(f_i,s_i)=0\}} &-&
\PbT{\bigcap_{i=1}^{m}\{\eta_T^{\textrm{edge}}(f_i^a,s_i)=0\}}\bigg|
\leq\nonumber \\ &\leq & \sum_{i=1}^m \PbT{\eta_T^\mathrm{edge}(g^a,s_i)\geq 1}.
\end{eqnarray}
The term \begin{equation}
\PbT{\bigcap_{i=1}^{m}\{\eta_T^{\textrm{edge}}(f_i^a,s_i)=0\}}
\end{equation}
converges to 
\begin{equation}
\Pb{\bigcap_{i=1}^{m}\{\eta^{\textrm{Airy}}(f_i^a/\kappa ,s_i \kappa/2)=0\}}
\end{equation}
which yields the right hand side of (\ref{eqBorder.1}) as $a\to\infty$.

The terms in the sum of the right hand side of (\ref{eqBorder.5}) are bounded by
\begin{equation}\label{eqBorder.8}
\PbT{\eta_T^\mathrm{edge}(g^a,s_i)\geq 1}\leq \ET{\eta_T^\mathrm{edge}(g^a,s_i)}
=\int_{a}^\infty \ET{\eta_T^\mathrm{edge}(r,s_i)}\dk{r}.
\end{equation}
From (\ref{eqEdge.18}),
\begin{equation}\label{eqBorder.9}
\ET{\eta_T^\mathrm{edge}(r,s_i)}\simeq\int_{0}^{\infty}\frac{1}{4\pi^2} J_1(-L)^2\dk{L}.
\end{equation}
$J_1(-L)$ is indeed a function of $r+L$, which asymptotics has been studied already for $r+L$ large, but bounded by $r+L\leq \e T^{2/3}$, with the result (\ref{eqEdge.83}). Therefore the integrals in (\ref{eqBorder.8}), (\ref{eqBorder.9}) converge for $r+L\leq \e T^{2/3}$.

Next consider $r+L > \e T^{2/3}$. Let $\wt{L}=(r+L)T^{-2/3}$. With the change of variable $u=\psi(k)$ and integrating twice by parts, we obtain
\begin{equation}
\abs{J_1(-L)}\leq T^{1/3} \frac{2 \psi(\pi)}{T^2} \max_{k\in[\psi(-\pi),\psi(\pi)]} \abs{\Dttt{k(u)}{u}}.
\end{equation}
Similarly as for (\ref{eqEdge.69b}) we have,
\begin{equation}
\max_{k\in[\psi(-\pi),\psi(\pi)]} \abs{\Dttt{k(u)}{u}} \leq  G_1 \wt{L}^{-2},
\end{equation}
for a suitable constant $G_1<\infty$, which yields
\begin{equation}
\abs{J_1(-L)}\leq G_2 (r+L)^{-2} T^{-1/3}
\end{equation}
for some constant $G_2<\infty$. Therefore the integrals in (\ref{eqBorder.8}), (\ref{eqBorder.9}) have a bound $G(a)$ uniform in $T$ which vanishes as $a\to\infty$

\end{proof}
\end{thm}

Probabilistically, it would be natural to lift Theorem~\ref{thmBorder} to
the weak convergence of path measures. The missing piece is the tightness for
the sequence of stochastic process $A_T(s)$. We have not attempted to fill this gap. The interested reader is referred to~\cite{Jo:DetProc}, where tightness for the edge scaling of the Aztec diamond is proved.

\appendix

\section{Fermionic correlations}
\subsection{Two-point function }\label{app1}
Let $\widehat{A}=\sum_{k,l\in \Z}{A_{k,l} a^*_k a_l}$ be the second quantization of
the one-particle matrix $A$. It is assumed that $e^{-A}$ is trace class and $\Det{1+e^{A}} \neq 0$ (see~\cite{ReedSimon}, Chap. XIII). We use the identities
\begin{equation}\label{eqApp.1}
e^{-\widehat{A}} a^*_i e^{\widehat{A}}=\sum_{j\in\Z}{a^*_j
[e^{-A}]_{j,i}},\quad
e^{-\widehat{A}} a_i e^{\widehat{A}}=\sum_{j\in\Z}{[e^{A}]_{i,j}a_j}.
\end{equation}
Then
\begin{eqnarray}
\ff{a^*_i a_j} &=&\frac{1}{Z}\Tr{e^{-\widehat{A}}a^*_i a_j} =
\sum_{n\in\Z}{\frac{1}{Z}\Tr{a^*_n [e^{-A}]_{n,i}e^{-\widehat{A}} a_j}}\\
&=& \sum_{n\in\Z}{[e^{-A}]_{n,i}(-\ff{a^*_n a_j}+\delta_{j,n})}
= [e^{-A}]_{j,i}-\sum_{n\in\Z}{\ff{a^*_n a_j} [e^{-A}]_{n,i}}, \nonumber
\end{eqnarray}
and
\begin{equation}
\sum_{n\in\Z}{\ff{a^*_n [\mathbbm{1}+e^{-A}]_{n,i} a_j}} = [e^{-A}]_{j,i}.
\end{equation}
Finally multiplying this expression by 
$\sum_{i \in \Z}{[(\mathbbm{1}+e^{-A})^{-1}]_{i,m}}$ we obtain
\begin{equation}
\ff{a^*_m a_j}=[(\mathbbm{1}+e^{A})^{-1}]_{j,m}.
\end{equation}

\subsection{Proof of (\ref{DetEq})-(\ref{DetEq2})}\label{app2}
We prove recursively that
\begin{equation}\label{eqApp.5}
\ff{a^*_{i_1}a_{j_1}\cdots a^*_{i_n}a_{j_n}}= \Det{R(i_k,j_l)}_{1\leq k,l \leq n},
\end{equation}
where
\begin{equation}
R(i_k,j_l)=\left\{\begin{array}{ll}\ff{a^*_{i_k}a_{j_l}}
& \textrm{ if }k \leq l,\\
-\ff{a_{j_l}a^*_{i_k}}
&\textrm{ if }k > l.\end{array}\right.
\end{equation}
Then, taking $i_k=j_k$ for all $k$, the result (\ref{DetEq})-(\ref{DetEq2}) is obtained.
For $n=1$ the formula holds by definition. Suppose the formula (\ref{eqApp.5})
has been established for some $n$, i.e.
\begin{equation} \label{eqn}
\ff{a^*_{i_1}a_{j_1}\cdots a^*_{i_n}a_{j_n}}=
\left\arrowvert\begin{array}{cccc}
\ff{a^*_{i_1}a_{j_1}} & \ff{a^*_{i_1}a_{j_2}} & \cdots & \ff{a^*_{i_1}a_{j_n}} \\
-\ff{a_{j_2}a^*_{i_1}} & \ff{a^*_{i_2}a_{j_2}} & \cdots & \ff{a^*_{i_2}a_{j_n}} \\
\vdots & \vdots & \ddots & \vdots \\
-\ff{a_{j_n}a^*_{i_1}} & -\ff{a_{j_n}a^*_{i_2}} & \cdots & \ff{a^*_{i_n}a_{j_n}}
\end{array}\right\arrowvert.
\end{equation}
We will need one more expression for $\ff{\cdots}$ such that in the first $k$
pairs the annihilation operator precedes the creation operator,
\begin{equation}\label{eqnk}
\begin{array}{l}
\ff{a_{j_1}a^*_{i_1}\cdots a_{j_k}a^*_{i_k}a^*_{i_{k+1}}a_{j_{k+1}} \ldots
a^*_{i_n}a_{j_n}}= \\[12pt]
=(-1)^k \left\arrowvert\begin{array}{cccccc}
-\ff{a_{j_1}a^*_{i_1}} & \cdots & \ff{a^*_{i_1}a_{j_k}} &
\ff{a^*_{i_1}a_{j_{k+1}}} &\cdots & \ff{a^*_{i_1}a_{j_n}} \\
\vdots & \ddots & \vdots & \vdots & \ddots & \vdots \\
-\ff{a_{j_k}a^*_{i_1}} & \cdots & -\ff{a_{j_k}a^*_{i_k}} &
\ff{a^*_{i_k}a_{j_{k+1}}} & \cdots & \ff{a^*_{i_k}a_{j_n}} \\
-\ff{a_{j_{k+1}}a^*_{i_1}} & \cdots & -\ff{a_{j_{k+1}}a^*_{i_k}}
& \ff{a^*_{i_{k+1}}a_{j_{k+1}}} & \cdots & \ff{a^*_{i_{k+1}}a_{j_n}}\\
\vdots & \ddots & \vdots & \vdots & \ddots & \vdots \\
-\ff{a_{j_n}a^*_{i_1}} & \cdots & -\ff{a_{j_n}a^*_{i_k}} &
-\ff{a_{j_n}a^*_{i_{k+1}}}& \cdots & \ff{a^*_{i_n}a_{j_n}}
\end{array}\right\arrowvert.
\end{array}
\end{equation}
Let us prove this formula. For $k=0$, it agrees with (\ref{eqn}). Suppose
it to be true for some $k$. Let us then prove that the formula (\ref{eqnk}) holds for $k+1$,
\begin{eqnarray}\label{eqA.9b}
& &\ff{a_{j_1}a^*_{i_1}\cdots a_{j_{k+1}}a^*_{i_{k+1}}a^*_{i_{k+2}}a_{j_{k+2}}
\ldots a^*_{i_n}a_{j_n}} = \nonumber\\
&&= -\ff{a_{j_1}a^*_{i_1}\cdots a_{j_k}a^*_{i_k}a^*_{i_{k+1}}a_{j_{k+1}} \ldots
a^*_{i_n}a_{j_n}} \\
& &\phantom{=}+ \delta_{i_{k+1},j_{k+1}} \ff{a_{j_1}a^*_{i_1}\cdots
a_{j_k}a^*_{i_k}a^*_{i_{k+2}}a_{j_{k+2}} \ldots a^*_{i_n}a_{j_n}}.\nonumber
\end{eqnarray}
Using the expression (\ref{eqnk}) and considering the expansion of the determinant in the $(k+1)^{\textrm{th}}$ column (or row), it is easy to see that (\ref{eqA.9b})  corresponds, up to a factor of $-1$, to the expression (\ref{eqnk}) but with the diagonal term $a^*_{i_{k+1}}a_{j_{k+1}}$ replaced by $-a_{j_{k+1}}a^*_{i_{k+1}}$. Therefore (\ref{eqnk}) holds for $k+1$, too.

Now we prove (\ref{eqn}) for $n+1$ by using (\ref{eqn}) for $n$ and (\ref{eqnk})
for $n$ and $k\leq n$,
\begin{eqnarray}
& &\ff{a^*_{q}a_{j_1}\cdots a^*_{i_{n+1}}a_{j_{n+1}}} = 
\frac{1}{Z}\Tr{e^{-\widehat{A}}a^*_{q}a_{j_1}\cdots a^*_{i_{n+1}}a_{j_{n+1}}}
\nonumber \\
& &= \sum_{m\in \Z}{\frac{1}{Z}[e^{-A}]_{m,q}\Tr{e^{-\widehat{A}}a_{j_1}\cdots a^*_{i_{n+1}}a_{j_{n+1}}a^*_m}}
\nonumber \\
& &= -\sum_{m\in \Z}{[e^{-A}]_{m,q}\,\ff{a^*_{m}a_{j_1}\cdots a^*_{i_{n+1}}a_{j_{n+1}}}}\\
& &\phantom{=} + \sum_{p=2}^{n+1}{[e^{-A}]_{j_p,q}\,\ff{a_{j_1}a^*_{i_2}\cdots a_{j_{p-1}}a^*_{i_p}a^*_{i_{p+1}}a_{j_{p+1}}\ldots
a^*_{i_{n+1}}a_{j_{n+1}}}} \nonumber \\
& &\phantom{=}+[e^{-A}]_{j_1,q}\,\ff{a^*_{i_2}a_{j_2}\cdots a^*_{i_{n+1}}a_{j_{n+1}}}.\nonumber
\end{eqnarray}
We take the term with the sum over $m\in \Z$ together with the first one and multiply the whole expression by 
$\sum_{q \in \Z}{[(\mathbbm{1}+e^{-A})^{-1}]_{q,i_1}}$ to obtain
\begin{eqnarray}
& &\ff{a^*_{i_1}a_{j_1}\cdots a^*_{i_{n+1}}a_{j_{n+1}}} = \ff{a^*_{i_1}a_{j_1}} \ff{a^*_{i_2}a_{j_2}\cdots a^*_{i_{n+1}}a_{j_{n+1}}}\\
& &+ \sum_{p=2}^{n+1}{\ff{a^*_{i_1}a_{j_p}}\ff{a_{j_1}a^*_{i_2}\cdots a_{j_{p-1}}a^*_{i_p}a^*_{i_{p+1}}a_{j_{p+1}}\ldots
a^*_{i_{n+1}}a_{j_{n+1}}}}.\nonumber
\end{eqnarray}
Using (\ref{eqn}) and (\ref{eqnk}) for $n$ terms we see that this last expression is nothing else than the expansion with respect to the first row of (\ref{eqn}) with $n$ substituted by $n+1$.

\section*{Acknowledgments}
We are most grateful to Michael Pr\"ahofer for his constant help and encouragement. The work of P.L. Ferrari was supported by the Swiss fellowship Sunburst-Fonds.

\end{document}